\documentclass[12pt]{article}
\usepackage{amsmath}
\usepackage{amssymb}
\begin{document}
%\begin{flushright}
%BHU-PHYS-CAS Preprint\\
%arXiv: 1005.5067 [hep-th]
%\end{flushright}
\begin{center}
{\bf {\large{Interacting 2D Field-Theoretic Model for Hodge Theory}}}

\vskip 2.5cm

{\sf  A. Tripathi$^{(a)}$, A. K. Rao$^{(a)}$, R. P. Malik$^{(a,b)}$}\\
$^{(a)}$ {\it Physics Department, Institute of Science,}\\
{\it Banaras Hindu University, Varanasi - 221 005, (U.P.), India}\\

\vskip 0.1cm

$^{(b)}$ {\it DST Centre for Interdisciplinary Mathematical Sciences,}\\
{\it Institute of Science, Banaras Hindu University, Varanasi - 221 005, India}\\
{\small {\sf {e-mails:  ankur1793@gmail.com; amit.akrao@gmail.com; 
  rpmalik1995@gmail.com}}}
\end{center}

\vskip 1.5 cm

\noindent
{\bf Abstract:}
We take up the St${\ddot u}$ckelberg-modified version of the  two (1+1)-dimensional (2D) Proca theory, in {\it interaction} 
with the Dirac fields, to study {\it its} 
various continuous and discrete symmetry transformations and show that {\it this} specific {\it interacting} 2D 
field-theoretic model provides a
tractable example for the Hodge theory because {\it its} symmetries (and corresponding conserved charges) provide 
the {\it physical}
realizations of the de Rham cohomological operators of differential geometry at the {\it algebraic} level. 
The physical 
state of {\it this} theory is chosen to be the harmonic state (of the Hodge decomposed state) in the quantum Hilbert space
which is annihilated by the {\it conserved} and {\it nilpotent} (anti-)BRST as well as (anti-)co-BRST charges.
A physical consequence of
{\it this} study is an observation that the 2D anomaly, at the {\it quantum} level, does {\it not} lead to any problem 
as far as the consistency and unitarity
of our present 2D theory is concerned. In other words, our present 2D field-theoretic model is amenable to 
particle interpretation 
{\it despite} the presence of the {\it local} {\it chiral} symmetry (which is associated with the nilpotent 
(anti-)co-BRST symmetry
transformations) {\it besides} the presence of the nilpotent (anti-)BRST symmetries 
(which are connected with the {\it local} {\it gauge} symmetry). The physicality condition with 
the (anti-)co-BRST
charges implies that the 2D anomaly term is {\it trivial} in our present theory. Hence, our 2D theory is
consistent, unitary and amenable to particle interpretation.


\vskip 0.5cm
\noindent
{\it {Keywords}}: St${\ddot u}$ckelberg {\it modified} 2D Proca theory; interacting Dirac fields; (anti-)BRST 
symmetries; (anti-)co-BRST 
symmetries; de Rahm cohomological operators; 
Hodge decomposition theorem; harmonic state as the physical state; physicality criteria; 2D anomaly and 
(anti-)co-BRST symmetries;
consistency and unitarity of our 2D theory

\newpage

\section {Introduction}

\noindent
Mathematics is the language of {\it all} the key branches of modern-day scientific pursuits. Its creative usefulness
and glorious presence are undeniable truths in the domain of theoretical physics. The applications and 
realizations of the mathematical ideas and concepts in the realm of theoretical physics have played very important
and decisive roles in the development of the {\it latter}. 
In this context, it is pertinent to point out that it has been possible to show, within the
framework of Becchi-Rouet-Stora-Tyutin (BRST) formalism [1-4], that the Abelian
$p$-form $(p = 1, 2, 3...)$ gauge theories in $D = 2p$ dimensions of spacetime provide the field-theoretic examples
of Hodge theory (see, e.g. [5-7] and references therein) where the discrete and continuous 
symmetries (and the conserved charges) have been able
to provide the physical realizations of the de Rham cohomological operators of differential geometry [8-12] 
at the {\it algebraic} level. Furthermore, it has been found that some of the
${\cal N} = 2$ supersymmetric (SUSY) quantum mechanical models are {\it also} examples of Hodge theory (see, e.g. [13-17])
at the {\it algebraic} level (without the central extension(s)). Whereas
the above Abelian $p$-form $(p = 1, 2, 3...)$ gauge theories are {\it massless} theories due to the presence of
{\it gauge} symmetries, the ${\cal N} = 2$ SUSY quantum mechanical models are {\it massive} and they respect {\it two} 
nilpotent SUSY transformations which are {\it not} absolutely anticommuting in nature. However, the (anti-)BRST
symmetries (corresponding to the gauge symmetry transformations of the Abelian $p$-form gauge theory) are
{\it nilpotent} of order two {\it but} absolutely anticommuting in nature. These properties are very {\it sacrosanct}.

Against the backdrop of the {\it above}, we have shown that the St$\ddot u$ckelberg-{\it modified} 2D Proca 
as well as the 4D Abelian 2-form gauge theories ({\it without} any interaction with 
{\it matter} fields) provide the {\it massive}
models of Hodge theories where the {\it mass} and the {\it quantum} (anti-)BRST as well as (anti-)co-BRST symmetries
co-exist {\it together} [18-20]. The central purpose of our
present investigation is to establish that the 2D {\it modified} Proca theory, in {\it interaction} with the Dirac fields, provides
a field-theoretic example of Hodge theory where the continuous symmetry transformations (and corresponding conserved
Noether charges) provide the physical realizations of the cohomological operators of differential geometry and the
existence of a couple of discrete symmetry transformations is at the heart of the physical realizations of the
Hodge duality $*$ operation of differential geometry. Thus, our present 2D theory is a {\it novel} example of 
a {\it massive} and {\it interacting} model of Hodge theory within the framework of BRST formalism. The existence of
the nilpotent (anti-)BRST and (anti-)co-BRST symmetries provides the physical realizations of the 
nilpotent (co-) exterior
derivatives of differential geometry. The (anti-)BRST symmetry transformations
correspond to the local {\it gauge} symmetry in the theory. On the other hand, it is the local {\it chiral}
symmetry transformation (in the massless limit of the Dirac fields) that leads to the existence of a proper set of (anti-)co-BRST
transformations in the theory.

The study of the field-theoretic models of Hodge theory, within the ambit of BRST formalism, is {\it physically} important
on the following grounds. First, it has been possible to show that the (non-)Abelian 1-form gauge theories, in two $(1 + 1)$-
dimensions of spacetime, provide a {\it new} [21] type of topological field theory (TFT) that captures a few key aspects
of the Witten-type TFT [22] and some salient features of the Schwarz-type TFT [23]. Second, 
we have been able to establish that the 4D Abelian
2-form and 6D Abelian 3-form gauge theories are tractable field-theoretic examples of Hodge theory and they provide
the models of {\it quasi}-TFTs where the topological invariants and their 
{\it perfect} recursion relations exist [5, 24, 25]. Third,
it has been established that the {\it massive} models of Hodge theory ({\it without} any interaction with matter fields)
allow the existence of fields with {\it negative} kinetic terms [18-20] which have been called as the ``phantom" fields in the
domain of cosmology. Such kinds of {\it exotic} fields have played important roles in the context of cyclic, bouncing
and self-accelerated models of Universe (see, e.g. [26-30]). In our earlier works [18-20], 
we have laid the emphasis on the relevance of such 
fields (with {\it negative} kinetic terms) in the context of providing a set of 
possible candidates for the dark matter and dark energy
within the framework of quantum field theory. Finally, in our present endeavor, we establish that the {\it massive}
and {\it interacting} 2D modified version of the Proca theory with the Dirac fields shows that the {\it vector} and 
{\it axial-vector} currents (and their corresponding charges) are conserved 
{\it together} because the 2D anomaly term is {\it trivial}
when we choose the physical state as the {\it harmonic} state (of the Hodge decomposed state in the {\it total} quantum Hilbert
space) which is annihilated by the conserved and nilpotent BRST and co-BRST charges {\it together}.

The following key factors have been at the heart of our present investigation. First, in the physical four
$(3 + 1)$-dimensions of QED with the Dirac fields, it has been well-established that the {\it vector} and {\it axial-vector}
currents can {\it not} be conserved {\it together} due to the presence of the anomaly term $(\sim \vec E \cdot \vec B)$
which is a pseudo-scalar constructed with the {\it polar-vector} electric field $(\vec E)$ and the
{\it axial-vector} magnetic field
$(\vec B)$ (see, e.g. [31] for details). However, in the case of 2D, there is presence of {\it only} the electric
field $(E)$ which becomes a pseudo-scalar and the {\it anomaly} term. We have been motivated to check  
the {\it status} of this anomaly {\it term} within the framework of BRST formalism. Our present endeavor is an attempt
in {\it that} direction. Second, we have been working on the idea of 
dual-BRST symmetry for quite sometime and we have shown [32, 33] that the 2D QED with the Dirac fields (interacting with
a {\it massless} photon) leads to the definition
of a conserved co-BRST charge which contains the pseudo-scalar electric field $(E)$. We have been curious to establish
a connection  between the co-BRST charge and the {\it anomaly} term $(E)$ in 2D. We have accomplished this goal in our present
endeavor. Finally, this is, for the {\it first} time, we have proven a {\it massive} as well as an {\it interacting}
gauge theory to be an example of the Hodge theory. We have already established the existence of the field-theoretic
models of Hodge theory with {\it mass} and {\it massless} gauge fields in our earlier works [18-20, 5-7]. Our present 2D
theory is {\it also} a perfect model of a {\it duality} invariant theory [34] because of the existence of the discrete
symmetry transformations [see, Eq. (10) below].

Our present {\it interacting} theory respects {\it six} continuous and a couple of discrete symmetry transformations {cf. Eq. (10) below].
Out of these, at the {\it most} economical level, {\it three} symmetries are very {\it fundamental}. These are the BRST, co-BRST and discrete symmetries.
The conserved and nilpotent BRST charge annihilates the physical (i.e. harmonic) state which implies that the operator forms  of
the {\it first-class} constraints {\it must} annihilate the physical state. The decisive presence of the discrete symmetry transformations  [cf. Eq. (10) below]
ensures that our theory is a {\it perfect} model (cf. Secs. 5, 6 below) of the {\it duality} invariant theory [34]. Hence, the dual versions
of the above first-class constraints must {\it also} annihilate the physical state. Thus, the annihilation of the physical state by the {\it first-class}
constraints and their {\it dual} is very {\it sacrosanct} and it must be respected at the {\it tree}  as well as at the {\it loop-level}
diagrams. These {\it dual} versions of constraints [cf. Eq. (47) below] emerge when we demand that the physical state must be annihilated
by the co-BRST charge which is nilpotent {\it but} conserved only at the {\it tree-level} diagrams because it is derived from the
co-BRST current [cf. Eq. (13) below] which is {\it also} conserved at the tree-level diagrams. However, it is the sanctity of the requirements
of a {\it perfect} duality-invariant theory that (i) the 2D anomaly term, and (ii) its invariance under the time-evolution,
must {\it also} annihilate the physical state of our theory [cf. Eq. (48) below]. This implies, primarily, that the
{\it vector} and {\it axial-vector} currents are found to be conserved {\it together} for our 2D interacting theory
with the Dirac fields. Hence, the 2D anomaly term becomes {\it trivial}.

The theoretical contents of our present endeavor are organized as follows. First, to set the notations and convention, 
we recapitulate the bare essentials of the (anti-)BRST symmetries of the D-dimensional Lagrangian density 
for the {\it modified} 
version of Proca (i.e. {\it massive} Abelian 1-form $A^{(1)} = d\,x^\mu\,A_\mu$) gauge theory where there is a coupling 
between the gauge field $A_\mu$ (with $\mu = 0, 1, 2...D-1)$ and the conserved current constructed {\it with} the 
fermionic Dirac fields in Sec. 2.
Our Sec. 3 is devoted to the discussion on the nilpotent (anti-)co-BRST symmetry transformations for the 
{\it same} [i.e. (anti-)BRST invariant]
Lagrangian density in two (1+1)-dimensions of spacetime. We derive the bosonic and ghost-scale symmetry
transformations in our Sec. 4. Our Sec. 5 deals with the derivation 
of the {\it algebras} obeyed by 
the {\it above} (anti-)BRST, (anti-)co-BRST, a {\it unique} bosonic and the ghost-scale
continuous symmetries {\it and} corresponding conserved charges.
In our Sec. 6, we elaborate on the connection between the {\it algebraic structures} of the above symmetries 
(and corresponding conserved charges)
{\it and} the de Rham cohomological operators of differential geometry. In Sec. 7, we dwell a bit on the 
irrelevance of the 2D anomaly term as far as our 
discussion on the importance of the (anti-)dual-BRST [i.e. (anti-)co-BRST] transformations (nd corresponding charges) is 
concerned. Finally, in Sec. 8, we summarize our results and lay emphasis on the {\it physical} importance
of our present work.

In our Appendices A, B and C, we perform some explicit computations which corroborate some of the claims 
that have been made in the main body of our text. \\

%In particular, our Appendix C is devoted to the proof that
%the {\it minimal} interaction term $(- e\,\bar\psi\,\gamma^\mu\,A_\mu\,\psi)$ remains invariant under the 
%discrete symmetry transformations [cf. Eq. (10) below] {\it even} in the component (i.e. matrix) forms of 
%$\bar\psi,\, \psi$ and $A_\mu$ in 2D.

\vskip 0.5cm
\noindent
{\it {Convention and Notations}}:
We adopt the background 2D Minkowskian {\it flat} spacetime manifold that is endowed with the metric tensor 
$\eta _{\mu\nu} =$  diag $(+1, -1)$ so that the dot product
$(P \cdot Q)$ between the two non-null vectors $P_\mu$ and $Q_\mu$ is defined as: 
$P \cdot Q = \eta_{\mu\nu}\,P^\mu\,Q^\nu = P_0\,Q_0 - P_i\,Q_i$
where the Greek indices $\mu,\, \nu,\, \lambda... = 0, 1$ correspond to the spacetime directions and $i, \,j, \,k,... = 1$ 
stand for the space direction.
The antisymmetric $(\varepsilon_{\mu\nu} = - \varepsilon_{\nu\mu} )$ Levi-Civita tensor $\varepsilon_{\mu\nu}$ 
is chosen such that $\varepsilon_{01} = 
 + \varepsilon^{10} = +1$ and $\varepsilon_{\mu\nu}\,\varepsilon^{\mu\nu} = -\,2!, \,$ $\varepsilon_{\mu\lambda}\,
\varepsilon^{\mu\nu} = -\,1!\,\delta^{\nu}_{\lambda}$, etc.
We denote the (anti-)BRST symmetry transformations by $s_{(a)b}$ and (anti-)co-BRST symmetry transformations 
by $s_{(a)d}$ in the whole body of our text. 
We take the convention of the left-derivative w.r.t. the fermionic fields ($\bar\psi,\, \psi,\, \bar C,\, C$) which 
anticommute among themselves {\it but} commute with {\it all} the bosonic fields of our theory. We choose the 
$ 2 \times 2 $ Dirac
 $\gamma$-matrices in terms of the traceless and hermitian $2 \times 2 $ Pauli-matrices as: 
$\gamma_0 = \sigma_1,\, \gamma_1 = i\, \sigma_2$ 
 which satisfy $\{\gamma_\mu, \,\gamma_\nu\} = 2\,\eta_{\mu\nu}$ and lead to the definition of $\gamma_5 = \gamma_0\,
\gamma_1 \equiv -\,\sigma_3$ which is the {\it diagonal} matrix [like the $\gamma_5$ matrix in the Weyl 
representation (see, e.g. [35] for details) of Dirac's matrices] and hermitian (i.e. $\gamma_5^\dagger = \gamma_5$). 
We {\it also} have: $(\gamma_5)^2 = I, \, \gamma_5\,\gamma_\mu = \varepsilon_{\mu\nu}\,\gamma^\nu,\, \gamma_5 
= - (1/2!)\,\varepsilon_{\mu\nu}\,\gamma^\mu\,\gamma^\nu$ and 
$\{\gamma_5, \gamma_\mu\} = 0$ where 
$\gamma_0 = \gamma^0$ and $\gamma^1 = - \,\gamma_1$ due to the choice of the signatures of our 2D flat metric tensor.

\vskip 0.5cm
\noindent
{\it Standard Definitions}:
We recapitulate the following definitions of the differential geometry [8-12] which will be useful in our discussions 
(as far as the connection between the {\it physical}
aspects and {\it mathematical} ideas of our present endeavor are concerned).\\

\noindent
(i) {\it The de Rham Cohomological Operators}: The set of {\it three} operators $(d,\, \delta ,\,\Delta )$ constitute 
the de Rham cohomological operators
(defined the D-dimensional compact manifold {\it without} a boundary) where $d = d\,x^\mu\,\partial_\mu$ 
(with $\mu = 0,\,1,\,2...D-1)$ is the 
exterior derivative, $\delta = \mp *\,d\,*$ is the co-exterior ({\it or} dual-exterior) derivative and 
$\Delta = (d + \delta)^2$ is the Laplacian operator.
Here $*$ is the Hodge duality operation on the above-mentioned compact manifold {\it without} a boundary. 
For the even dimensions of spacetime 
$\delta = -\,*\,d\,*$. These operators obey the algebra: $d^2 = \delta^2 = 0, \, \Delta  = 
(d + \delta)^2 = d\,\delta + \delta \,d = \{d,\,\delta\}$,\,
$[\Delta,\, d] = 0,\, [\Delta,\,\delta] =  0$ which is popularly known as the Hodge algebra (that is {\it not} a Lie algebra).\\ 

\noindent
(ii) {\it The Hodge Decomposition Theorem}: 
On a compact manifold {\it without} a boundary, any arbitrary form $f_n$ (of {\it non-zero} degree $n$) can be
{\it uniquely} written as the {\it sum} of the harmonic form $(h_n)$,
an exact form $(d\,g_{n-1})$ and  a co-exact form $(\delta\,k_{n+1})$ as: $f_n = h_n + d\,g_{n-1} + \delta\,k_{n+1}$ 
where the harmonic form $h_n$
satisfies: $d\, h_n  = \delta \,h_n = 0$. In other words, we have $\Delta \, h_n = 0$. Thus, the harmonic 
form $(h_n)$ is {\it closed} and 
{\it co-closed} together.\\

\section {Preliminaries: (Anti-)BRST Symmetry Invariant Lagrangian Density in Any Arbitrary Dimension}

We begin with the following D-dimensional (anti-)BRST invariant {\it interacting} Lagrangian density in the 
't Hooft gauge (see, e.g. [36] for details)
\begin{eqnarray}
{\cal L}_{B} &=& - \frac{1}{4}\,F_{\mu\nu}\,F^{\mu\nu} + \frac{m^2}{2}\,{A_\mu}{A^\mu} + \frac{1}{2}\,
\partial_\mu\,\phi\,\partial^\mu\,\phi - m\,A_\mu\,\partial^\mu\,\phi + \bar \psi\,(i\,\gamma^\mu\,\partial_\mu -
m')\,\psi \nonumber\\
&-& e\,\bar \psi\,\gamma^\mu\,A_\mu\,\psi + B\,(\partial \cdot A + m\,\phi) + \frac{1}{2}\,B^2 - i\,\partial_\mu\,\bar C\,
\partial^\mu\,C + i\,m^2\,\bar C\,C.
\end{eqnarray}
where the kinetic term for the {\it massive} photon has its origin in the exterior derivative 
$(d = d\,x^\mu\,\partial_\mu$ with $d^2 = 0)$
of the differential geometry [8-12] because we note that $d\,A^{(1)} = \frac{1}{2!}\,(d\,x^\mu\wedge d\,x^\nu)\,F_{\mu\nu}$.
Here $A^{(1)} = d\,x^\mu\,A_{\mu}$ (with $\mu = 0,\,1,\,2...D-1)$ defines the {\it massive} vector field 
$(A_\mu)$ with rest mass $m$ and 
$F_{\mu\nu} = \partial_\mu\,A_\nu - \partial_\nu\,A_\mu$ is the field strength tensor. The {\it pure} scalar field $\phi$ 
has been invoked because of 
St${\ddot u}$ckelberg's formalism and it has {\it positive} kinetic term. The fermionic 
$(\psi^2 = \bar\psi^2 = 0,\, \psi\,\bar\psi + \bar\psi\,\psi = 0)$
 Dirac fields (with the rest mass $m'$) have {\it interaction} with the {\it massive} gauge field $A_\mu$ through {\it minimal} 
interaction $(-\,e\,\bar \psi\,\gamma^\mu\,\psi\,A_\mu)$ coupling.
 The gauge-fixing term for the gauge field owes its origin to the exterior derivative $(\delta = - *\,d\,*)$ 
because $\delta\,A^{(1)} = (\partial \cdot A)$
 and we have added the $m\,\phi$ term to it on the dimensional ground. In our theory, $B$ is the Nakanishi-Lautrup 
auxiliary field and the fermionic
 $(C^2 = \bar C^2 = 0,\, C\,\bar C + \bar C \,C = 0)$ (anti-)ghost fields $(\bar C)\,C$ have been invoked 
to preserve the {\it unitarity} in the theory
at any order of perturbative computations.
The above Lagrangian density $({\cal L}_B)$ respects the following off-shell nilpotent $[s_{(a)b}^2 = 0]$
and absolutely anticommuting $(s_b\,s_{ab} + s_{ab}\,s_b = 0)$ (anti-)BRST transformations $[s_{(a)b}]$,
namely;
\begin{eqnarray} 
&& s_{ab}\,A_\mu = \partial_\mu\,\bar C, \qquad s_{ab}\,\phi = m\,\bar C, \qquad 
s_{ab}\,\psi = - i\,e\,\bar C\,\psi, \qquad s_{ab}\,\bar \psi = - i\,e\,\bar \psi\, \bar C, \nonumber\\
&& s_{ab}\,\bar C = 0, \qquad s_{ab}\,C = - i\,B, \qquad s_{ab}\,B = 0, \nonumber\\
&& s_b\,A_\mu = \partial_\mu\,C, \qquad s_b\,\phi = m\,C, \qquad s_b\,\psi = - i\,e\,C\,\psi, \qquad s_b\,\bar \psi 
= - i\,e\,\bar \psi\,C, \nonumber\\
&& s_b\,C = 0, \qquad s_b\,\bar C = i\,B, \qquad s_b\,B = 0,
\end{eqnarray} 
because the Lagrangian density $({\cal L}_B)$ transforms to
\begin{eqnarray}
s_{ab}\,{\cal L}_B = \partial_\mu\,[B\,\partial^\mu\,\bar C], \qquad 
s_b\,{\cal L}_{B} = \partial_\mu\,[B\,\partial^\mu\,C],
\end{eqnarray}
thereby rendering the action integral $S = \int d^D\,x\,{\cal L}_B$ invariant [i.e. $s_{(a)b}\,S = 0$]
for the physical fields that vanish-off as $x \rightarrow \pm\,\infty$ due to Gauss's divergence theorem.
As a consequence of the celebrated Noether's theorem, we have the conserved currents $[J^\mu_{(a)b}]$ corresponding to the
off-shell nilpotent, absolutely anticommuting, infinitesimal and continuous (anti-)BRST symmetry transformations
$s_{(a)b}$ as:
\begin{eqnarray}
J^\mu_{(ab)} &=& B\,\partial^\mu\,\bar C + m\,\bar C\,\partial^\mu\,\phi - F^{\mu\nu}\,\partial_\nu\,\bar C 
- m^2\,A^\mu\,\bar C - e\,\bar\psi\,\gamma^\mu\,\bar C\,\psi, \nonumber\\
J^\mu_{(b)} &=& B\,\partial^\mu\,C + m\,C\,\partial^\mu\,\phi - F^{\mu\nu}\,\partial_\nu\,C - m^2\,A^\mu\,C 
- e\,\bar \psi\,\gamma^\mu\,C\,\psi.
\end{eqnarray}  
The conservation law $[\partial_\mu\,J^\mu_{(a)b} = 0]$ can be proven readily by using the following
Euler-Lagrange (EL) equations of motion (EOMs) from the Lagrangian density $({\cal L}_B)$, namely;   
\begin{eqnarray}
&& \partial_\mu\,F^{\mu\nu} + m^2\,A^\nu - \partial^\nu\,B - m\,\partial^\nu\,\phi - e\,\bar \psi\,\gamma^\nu\,\psi = 0,
\qquad \Box\,\phi - m\,(\partial \cdot A) - m\,B = 0, \nonumber\\
&& i\,(\partial_\mu\,\bar \psi)\,\gamma^\mu + m'\,\bar \psi + e\,\bar \psi\,\gamma^\mu\,A_\mu = 0, \qquad 
i\,\gamma^\mu\,\partial_\mu\,\psi - m'\,\psi - e\,\gamma^\mu\,A_\mu\,\psi = 0, \nonumber\\
&& (\Box + m^2)\,\bar C = 0, \qquad (\Box + m^2)\,C = 0, \qquad B + \partial \cdot A + m\,\phi = 0. 
\end{eqnarray}
The conserved (anti-)BRST charges $(Q_{ab} = \int d^{D - 1}\,x\,J^0_{(ab)}, \, Q_{b} = \int d^{D - 1}\,x\,J^0_{(b)})$
are 
\begin{eqnarray}
Q_{ab} &=& \int d^{D - 1}\,x \, \Big[B\,\dot{\bar C} + m\,\dot \phi\,\bar C - F^{0i}\,\partial_i\,\bar C 
- m^2\,A^0\,\bar C - e\,\bar \psi\,\gamma^0\,\bar C\,\psi\Big] \nonumber\\ 
&\equiv& \int d^{D - 1}\,x \,\big(B\,\dot{\bar C} - \dot B\,\bar C \big), \nonumber\\
Q_b &=& \int d^{D - 1}\,x \, \Big[B\,\dot C + m\,\dot \phi\,C - F^{0i}\,\partial_i\,C - m^2\,A^0\,C 
- e\,\bar \psi\,\gamma^0\,C\,\psi\Big] \nonumber\\ 
&\equiv& \int d^{D - 1}\,x \,\big(B\,\dot C - \dot B\,C \big),
\end{eqnarray}
where the {\it final} expressions for the concise form of the conserved charges have been obtained 
due to the following EL-EOM that has been chosen from (5), namely;
\begin{eqnarray}
\partial_i\,F^{0i} = m^2\,A^0 - \dot B - m\,\dot \phi - e\,\bar \psi\,\gamma^0\,\psi.
\end{eqnarray}  
In other words, first of all, we have expressed $(- F^{0i}\,\partial_i\,\bar C)$ and $(- F^{0i}\,\partial_i\,C)$
as $[(\partial_i\,F^{0i})\,\bar C - \partial_i\,(F^{0i}\,\bar C)]$ and  $[(\partial_i\,F^{0i})\,C - 
\partial_i\,(F^{0i}\,C)]$, respectively, and applied the Gauss divergence theorem to drop the total {\it space}
derivative terms.  
It is straightforward to note that the conserved (anti-)BRST charges are off-shell nilpotent $[Q_{(a)b}^2 = 0]$
of order two and absolutely anticommuting $(Q_b\,Q_{ab} + Q_{ab}\,Q_b = 0)$ in nature due to the following 
observations in terms of the conserved charges (6) and the nilpotent (anti-)BRST transformations (2), namely;
\begin{eqnarray}
 s_b\,Q_b = - i\,\{Q_b, Q_b\} = 0 \qquad &\Rightarrow& \qquad Q_b^2 = 0, \nonumber\\
 s_{ab}\,Q_{ab} = - i\,\{Q_{ab}, Q_{ab}\} = 0 \qquad &\Rightarrow& \qquad Q_{ab}^2 = 0, \nonumber\\
 s_{b}\,Q_{ab} = - i\,\{Q_{ab}, Q_{b}\} = 0 \qquad &\Rightarrow& \qquad \{Q_{ab}, Q_b\} = 0, \nonumber\\
 s_{ab}\,Q_{b} = - i\,\{Q_{b}, Q_{ab}\} = 0 \qquad &\Rightarrow& \qquad \{Q_{b}, Q_{ab}\} = 0,
\end{eqnarray}
where the l.h.s. of the above relationships have been computed by {\it directly} applying the (anti-)BRST
symmetry transformations [cf. Eq. (2)] {\it on} the expressions for the nilpotent (anti-)BRST charges [cf. Eq. (6)]
where the {\it basic} principle behind the continuous symmetry transformations and {\it their} generators (as the 
conserved {\it Noether} charges) has been exploited to establish the off-shell nilpotency 
$([Q_{(a)b}]^2 = 0)$ and absolute anticommutativity
$(\{Q_b, Q_{ab}\} = 0)$
of the conserved ${\dot Q}_{(a)b} = 0$ (anti-)BRST charges $[Q_{(a)b}]$.

We wrap-up this section with the following crucial and decisive remarks. First, we note that the kinetic term of the
gauge field $(- \frac{1}{4}\,F_{\mu\nu}\,F^{\mu\nu})$ remains invariant (i.e. $s_{(a)b}\,F_{\mu\nu} = 0$) under the (anti-)BRST
symmetry transformations $[s_{(a)b}]$. Second, as pointed out earlier, the kinetic term has its origin in the exterior derivative
of differential geometry. Third, we observe that the ghost number is {\it increased} by {\it one} by the 
BRST symmetry transformation
when it applies on a generic field (e.g. $s_b\,A_\mu = \partial_\mu\,C,\, s_b\,\bar C = i\,B$, etc.) 
in contrast to the anti-BRST symmetry
transformations that {\it decreases} (e.g. $s_{ab}\,A_\mu = \partial_\mu\,\bar C,\, s_{ab}\,C = - i\,B$, etc.) 
the ghost number by {\it one}.
Fourth, the absolute anticommutativity (i.e. $Q_b\,Q_{ab} + Q_{ab}\,Q_b = 0,\, s_b\,s_{ab} + s_{ab}\,s_b = 0$) 
of the (anti-)BRST charges
and the continuous fermionic $[s_{(a)b}^2 = 0]$ symmetry transformations ensure that the (anti-)BRST symmetries 
(and their corresponding
conserved charges) can {\it not} be {\it together} identified with the {\it exterior} derivative of differential
 geometry. Fifth,
the absolute anticommutativity property ensures that the nilpotent (anti-)BRST symmetry transformations are 
{\it different} from the ${\cal N} = 2$ 
SUSY symmetry transformations which are nilpotent {\it but} not absolutely anticommuting in nature. Sixth, it 
is evident that, for the 
identification of the {\it co-exterior} derivative, we have to invoke and discuss {\it another} kind of nilpotent 
symmetries for our theory.

\section{(Anti-)Dual-BRST Symmetries: Modified Interacting 2D Proca Theory with Dirac Fields}

As we have modified the {\it mass} term $(\frac{m^2}{2}\,A_\mu\,A^\mu)$ by exploiting the theoretical
tricks of the St$\ddot u$ckelberg formalism [36], in exactly similar fashion, we can {\it modify} the 2D kinetic
term $(- \frac{1}{4}\,F_{\mu\nu}\,F^{\mu\nu} = \frac{1}{2}\,E^2)$ of the gauge field by invoking a 
pseudo-scalar field $\tilde\phi$ along with a
linearizing Nakanishi-Lautrup type auxiliary field ${\cal B}$ as (see, e.g. [19] for details):
\begin{eqnarray}
{\cal L}_{\cal B} &=& {\cal B}\,(E - m\,\tilde\phi) - \frac{1}{2}\,{\cal B}^2 + m\,E\,\tilde\phi 
- \frac{1}{2}\,\partial_\mu\,\tilde\phi\,\partial^\mu\,\tilde\phi + \frac{m^2}{2}\,{A_\mu}{A^\mu} + \frac{1}{2}\,
\partial_\mu\,\phi\,\partial^\mu\,\phi \nonumber\\
&-& m\,A_\mu\,\partial^\mu\,\phi + \bar \psi\,(i\,\gamma^\mu\,\partial_\mu -
m')\,\psi - e\,\bar \psi\,\gamma^\mu\,A_\mu\,\psi + B\,(\partial \cdot A + m\,\phi) + \frac{1}{2}\,B^2 \nonumber\\ 
&-& i\,\partial_\mu\,\bar C\,
\partial^\mu\,C + i\,m^2\,\bar C\,C.
\end{eqnarray}
It should be noted that the kinetic term $- \frac{1}{2}\,\partial_\mu\,\tilde\phi\,\partial^\mu\,\tilde\phi$ 
for the 
pseudo-scalar field $\tilde\phi$ carries a {\it negative} sign with it. This is due to the fact that we have the 
following {\it discrete}
symmetry in our theory (modulo total derivatives) provided we have the {\it positive} 
and {\it negative} signs for the kinetic terms for 
the pure scalar $(\phi)$ and pseudo-scalar
$(\tilde\phi)$ fields, respectively, namely;
\begin{eqnarray}
&& A_\mu \to \mp\, i\,\varepsilon_{\mu\nu} A^\nu, \quad \phi \to \mp\, i\, \tilde \phi, 
\quad \tilde \phi \to \mp \,i \, \phi,
\quad C \to \pm\, i\, \bar C, \quad \bar C \to \pm\, i\, C, \nonumber\\
&& {\cal B} \to \mp\; i\, B, \quad B \to \mp\; i\, {\cal B},  \quad \psi \to \psi, 
\quad \bar\psi \to \bar\psi,
\quad e \to \pm\; i\,e\,\gamma_5, \nonumber\\
&&(\partial \cdot A) \to \pm\; i\, E,    \quad  E \to \pm \;i \,(\partial \cdot A), \quad 
- m\,A_\mu\,\partial^\mu\,\phi \to + m\,E\,\tilde\phi.
\end{eqnarray}
We would like to mention that the mass term $(\frac{m^2}{2}\,A_\mu\,A^\mu)$ for the gauge field $(A_\mu)$ 
remains invariant under
$(A_\mu \to \mp\, i\,\varepsilon_{\mu\nu} A^\nu)$. Hence, the {\it masses} associated with $\phi$ and 
$\tilde\phi$ fields are the {\it same}
[{\it otherwise} the discrete symmetry (10) will be {\it violated}]. We shall see {\it later} that the 
above {\it two} discrete symmetry transformations (10) are very
{\it important} for our whole discussion because they provide the physical realizations of the Hodge duality 
$*$ operation of differential
geometry [8-12] as the analogue of relationship: $\delta = \pm\,* \,d \,*$ in terms of the off-shell 
nilpotent (anti-)BRST and 
(anti-)co-BRST transformations (cf. Sec. 5).

It is very interesting to point out that the 2D Lagrangian density (9) respects the following (anti-)co-BRST
symmetry transformations $[s_{(a)d}]$, namely;
\begin{eqnarray} 
&& s_{ad}\,A_\mu = - \varepsilon_{\mu\nu}\,\partial^\nu\,C, \quad s_{ad}\,\tilde\phi = - m\,C, \quad 
s_{ad}\,\psi =  i\,e\,C\,\gamma_5\,\psi, \quad s_{ad}\,\bar \psi =  - i\,e\,\bar\psi\,\gamma_5\,C, \nonumber\\
&& s_{ad}\,\bar C = i\,{\cal B}, \quad s_{ad}\,C = 0, \quad s_{ad}\,B = s_{ad}\,{\cal B} = s_{ad}\,\phi = 0, \quad 
s_{ad}\,E = \Box C, \nonumber\\
&& s_d\,A_\mu = - \varepsilon_{\mu\nu}\,\partial^\nu\,\bar C, \quad s_d\,\tilde\phi = - m\,\bar C, \quad 
s_d\,\psi =  i\,e\,\bar C\,\gamma_5\,\psi, \quad s_d\,\bar \psi =  - i\,e\,\bar\psi\,\gamma_5\,\bar C, \nonumber\\
&& s_d\,\bar C = 0, \quad s_d\,C = - i\,{\cal B}, \quad s_d\,B = s_d\,{\cal B} = s_d\,\phi = 0, \quad 
s_d\,E = \Box \bar C,
\end{eqnarray} 
because [for the massless $(m' = 0)$ fermions], we have the following
\begin{eqnarray} 
&& s_{ad}\,{\cal L}_{\cal B} = \partial_\mu\,\Big[{\cal B}\,\partial^\mu\,C + m\,\varepsilon^{\mu\nu}\,(m\,C\,A_\nu 
+ \phi\,\partial_\nu\,C) + m\,\tilde\phi\,\partial^\mu\,C \Big] , \nonumber\\
&& s_d\,{\cal L}_{\cal B} = \partial_\mu\,\Big[{\cal B}\,\partial^\mu\,\bar C + m\,\varepsilon^{\mu\nu}\,(m\,\bar C\,A_\nu 
+ \phi\,\partial_\nu\,\bar C) + m\,\tilde\phi\,\partial^\mu\,\bar C \Big],
\end{eqnarray}
which render the action integral $S = \int d^2\,x\,{\cal L}_{\cal B}$ {\it invariant} for the physical fields
that vanish-off as $x \rightarrow \pm\,\infty$ due to Gauss's divergence theorem. According to Noether's theorem,
the above continuous symmetry transformations lead to the derivation of the Noether {\it conserved} currents 
$[J^\mu_{(r)}, r = d, ad]$ at the classical (i.e. {\it tree} Feynman diagram) level as:
\begin{eqnarray}
 J^\mu_{(ad)} &=& {\cal B}\,\partial^\mu\,C + m\,C\,\partial^\mu\,\tilde\phi \nonumber\\
&-& \varepsilon^{\mu\nu}\big[B\,(\partial_\nu\,C) + m^2\,C\,A_\nu + m\,\phi\,(\partial_\nu\,C) \big] 
+ e\,\bar\psi\,\gamma^\mu\gamma_5\,C\,\psi, \nonumber\\
 J^\mu_{(d)} &=& {\cal B}\,\partial^\mu\,\bar C + m\,\bar C\,\partial^\mu\,\tilde\phi \nonumber\\
&-& \varepsilon^{\mu\nu}\big[B\,(\partial_\nu\,\bar C) + m^2\,\bar C\,A_\nu 
+ m\,\phi\,(\partial_\nu\,\bar C) \big]  
+ e\,\bar\psi\,\gamma^\mu\gamma_5\,\bar C\,\psi.
\end{eqnarray}
The conservation law for the above currents requires that $\partial_\mu J^\mu_{(5)} = \partial_\mu (\bar \psi\, \gamma^\mu\, \gamma_5\, \psi) = 0 $
which is true only at the tree-level Feynman diagram (when the massless limit of the Dirac fields are taken into account). These tree-level
conserved currents lead to the definition of the conserved $(\dot Q_{(r)} = 0$ with $r = d, ad)$
charges $Q_{(r)} = \int d\,x\,J_r^0$ (with $r = d, ad$)
as:
\begin{eqnarray}
Q_{ad} &=& \int d\,x \, \Big[{\cal B}\,\dot{C} + m\,\dot{\tilde\phi}\,C +
B\,\partial_1\,C + m^2\,A_1\,C + m\,\phi\,\partial_1\,C  + e\,
\bar \psi\,\gamma^0\,\gamma_5\,C\,\psi\Big], \nonumber\\
Q_{d} &=& \int d\,x \, \Big[{\cal B}\,\dot{\bar C} + m\,\dot{\tilde\phi}\,\bar C +
B\,\partial_1\,\bar C + m^2\,A_1\,\bar C + m\,\phi\,\partial_1\,\bar C  \nonumber\\
&+& e\,\bar \psi\,\gamma^0\,\gamma_5\,\bar C\,\psi\Big].
\end{eqnarray}
It is worthwhile to mention here that the conservation law $(\partial_\mu\,J^\mu_{(r)} = 0, \, r = d, ad)$ for
the Noether conserved currents (13) and conserved $({\dot Q}_{(r)} = 0,\, r = d, ad)$ charges (14) can be readily
proven, at the classical (i.e. {\it tree} Feynman diagram) level, by using the following EL-EOMs that are derived from the Lagrangian density (9):
\begin{eqnarray}
&& \varepsilon^{\mu\nu}\,\big[\partial_\nu\,{\cal B} + m\,\partial_\nu\,\tilde\phi \big] = m^2\,A^\mu - \partial^\mu\,B
- m\,\partial^\mu\,\phi - e\,\bar \psi\,\gamma^\mu\,\psi, \nonumber\\
&& \Box\,\tilde\phi + m\,E - m\,{\cal B} = 0, \qquad \Box\,\phi - m\,(\partial \cdot A) - m\,B = 0, \nonumber\\
&& i\,(\partial_\mu\,\bar \psi)\,\gamma^\mu + m'\,\bar \psi + e\,\bar \psi\,\gamma^\mu\,A_\mu = 0, \qquad 
i\,\gamma^\mu\,\partial_\mu\,\psi - m'\,\psi - e\,\gamma^\mu\,A_\mu\,\psi = 0, \nonumber\\
&& (\Box + m^2)\,\bar C = 0, \qquad (\Box + m^2)\,C = 0, \qquad B = - \,(\partial \cdot A + m\,\phi) \nonumber\\ 
&& {\cal B} = (E - m\,\tilde\phi) \equiv - (\varepsilon^{\mu\nu}\,\partial_\mu\,A_\nu + m\,\tilde\phi). 
\end{eqnarray}
A close look and keen observation of these equations imply that we have: $(\Box + m^2)\,\phi = 0,\, 
(\Box + m^2)\,\tilde\phi = 0,\, (\Box + m^2)\,B = 0,\, (\Box + m^2)\,{\cal B} = 0$. The last entry [i.e. 
$(\Box + m^2)\,{\cal B} = 0$] is {\it true} only at the {\it classical} level (i.e. {\it tree-level} Feynman diagrams) 
in the {\it massless} limit $(m' = 0)$
of the Dirac fields. This is due to the fact that we use: $\partial_\mu\,(\bar\psi\,\gamma^\mu\,\gamma_5\,\psi) = 0$.
We shall offer more comments on it in Sec. 7 where the {\it 2D anomaly term} will be discussed. 
It is also worthwhile to mention that in the derivation of 
$(\Box + m^2)\,B = 0$, we have used the conservation of vector current (i.e. $\partial_\mu\,J^\mu_{(m)} = e\,\partial_\mu\,
[\bar\psi\,\gamma^\mu\,\psi] = 0$) that is {\it true} at the {\it classical} as well as {\it quantum} level.  
Furthermore, we can derive the {\it simpler} forms of the conserved charges $Q_r \, (r = d, ad)$ using the 
Gauss's divergence theorem and the following
EL-EOM that has been taken from the top entry of Eq. (15), namely;
\begin{eqnarray}
\dot{\cal B} =   \partial_1\,B + m\,(\partial_1\,\phi) + e\,
\bar\psi\,\gamma_1\,\psi - m^2\,A_1  - m\,\dot{\tilde\phi},
\end{eqnarray}
which leads to the following concise expressions 
\begin{eqnarray}
Q_{ad}^{(1)} &=& \int d\,x \,\big({\cal B}\,\dot{C} - \dot{\cal B}\,C \big), \nonumber\\
Q_{d}^{(1)} &=& \int d\,x \,\big({\cal B}\,\dot{\bar C} - \dot{\cal B}\,\bar C \big).
\end{eqnarray}
It is now crystal clear, from the (anti-)co-BRST symmetry transformations (11) and the above
concise expressions for the conserved (anti-)co-BRST charges (17), that we have the following
explicit relationships, namely;
\begin{eqnarray}
s_d\,Q_d^{(1)} = - i\,\{Q_d^{(1)}, Q_d^{(1)}\} = 0 \quad &\Rightarrow& \quad [Q_d^{(1)}]^2 = 0, \nonumber\\
s_{ad}\,Q_{ad}^{(1)} = - i\,\{Q_{ad}^{(1)}, Q_{ad}^{(1)}\} = 0 \quad &\Rightarrow& \quad [Q_{ad}^{(1)}]^2 = 0, \nonumber\\
s_d\,Q_{ad}^{(1)} = - i\,\{Q_{ad}^{(1)}, Q_d^{(1)}\} = 0 \quad &\Rightarrow& \quad Q_{ad}^{(1)}\,Q_d^{(1)} +
Q_d^{(1)}\,Q_{ad}^{(1)} = 0, \nonumber\\
s_{ad}\,Q_{d}^{(1)} = - i\,\{Q_{d}^{(1)}, Q_{ad}^{(1)}\} = 0 \quad &\Rightarrow& \quad Q_{d}^{(1)}\,Q_{ad}^{(1)} +
Q_{ad}^{(1)}\,Q_{d}^{(1)} = 0.
\end{eqnarray}
The above equations capture the off-shell nilpotency (i.e. $[Q_{(r)}^{(1)}]^2 = 0,\,r = d, ad$) and absolute
anticommutativity (i.e. $Q_d^{(1)}\,Q_{ad}^{(1)} + Q_{ad}^{(1)}\,Q_{d}^{(1)} = 0$) of the (anti-)co-BRST charges
where we have exploited the theoretical strength of the continuous symmetry transformations and {\it their}
generators as the conserved Noether charges. It is straightforward to compute the {\it l.h.s.} of Eq. (18)
by {\it directly} applying the (anti-)co-BRST symmetry transformations (11) on the concise expressions for the
conserved $({\dot Q}_{(a)d}^{(1)} = 0)$ (anti-)co-BRST charges that have been quoted in Eq. (17) 
to check the sanctity of Eq. (18).

We wrap-up this section with the following key and decisive remarks. First of all, we note that the (anti-)co-BRST
symmetry transformations (11) are off-shell nilpotent (i.e. $s_{(a)d}^2 = 0$) of  
order two and they are absolutely anticommuting (i.e. $s_d\,s_{ad} + s_{ad}\,s_d = 0$) in nature. Second,
we observe that the {\it total} gauge-fixing term [i.e. $(\partial.A + m\,\phi)$] remains {\it invariant}
under the (anti-)co-BRST symmetry transformations [i.e. $s_{(a)d}\,(\partial.A) = 0, \, s_{(a)d}\,\phi = 0$]. Third, 
as already mentioned earlier, the gauge-fixing term $(\partial.A)$ of the gauge field $(A_\mu)$ 
owes its origin to the co-exterior derivative because $\delta\,A^{(1)} = -\,*\,d\,*\,(d\,x^\mu\,A_\mu) 
= (\partial.A)$. In the total gauge-fixing term $(\partial.A + m\,\phi)$, it is evident that the term 
$m\,\phi$ has been added on the dimensional ground (in the natural units where $\hbar = c = 1$). Fourth, to be
precise, we have {\it four} forms of Lagrangian densities for our 2D theory as discussed thoroughly in our earlier
work [19]. However, we have chosen {\it only} one for the sake of bravity in Eq. (9). We have provided
a synopsis of the varieties of Lagrangian densities in our Appendix A. Fifth, 
the absolute anticommutativity $(s_d\,s_{ad} + s_{ad}\,s_d = 0)$ of the
(anti-)co-BRST symmetry transformations (11) (and their intimate connection with the invariance of the
gauge-fixing term 
owing its origin to the co-exterior derivative $\delta = -\,*\,d\,*$) demonstrates that only {\it one} of them 
can be identified with $\delta$. Sixth, we note that the co-BRST symmetry transformations decrease the
ghost number by one (i.e. $s_d\,A_\mu = - \varepsilon_{\mu\nu}\,\partial^\nu\,\bar C,\, s_d\,C = - i\,{\cal B}$, etc.).
On the contrary, the anti-co-BRST symmetry transformations increase (i.e. $s_{ad}\,A_\mu = 
- \varepsilon_{\mu\nu}\,\partial^\nu\,C,\, s_{ad}\,\bar C =  i\,{\cal B}$, etc.) the ghost number by {\it one}
for the fields on which they act. Finally, we note that the conservation of the currents in (13)
and charges [cf. Eqs. (14), (17)] is valid only at the classical (i.e. tree Feynman diagram) level. We shall
discuss more on {\it this} issue in our Sec. 7 to show that the conservation law {\it persists} at the loop-level, too.\\

\section{Unique Bosonic and Ghost-Scale Continuous Symmetry Transformations: Conserved Charges}

It is evident that, so far, we have discussed {\it four} fermionic [i.e. (anti-)BRST and (anti-)co-BRST] 
symmetry transformations for our theory. We have also noted that $s_b\,s_{ab} + s_{ab}\,s_b = 0$ and 
$s_d\,s_{ad} + s_{ad}\,s_d = 0$ due to the absolute anticommutativity properties of the (anti-)BRST and
(anti-)co-BRST symmetry transformations. With the above observations as inputs, we note that we have the 
following bosonic symmetry transformations $(s_w, s_{\bar w})$, namely;
\begin{eqnarray}        
s_w = \{s_b, s_d\}, \qquad s_{\bar w} = \{s_{ad}, s_{ab}\},
\end{eqnarray}
that are constructed with the existing {\it four} fermionic symmetry transformations $s_{(a)b}$ and $s_{(a)d}$
of our theory. These explicit symmetry transformations are 
\begin{eqnarray}
&& s_w\,A_\mu = - i\,\varepsilon_{\mu\nu}\,\partial^\nu\,B - i\,\partial_\mu\,{\cal B}, \qquad 
s_w\,\phi = - i\,m\,{\cal B}, \qquad s_w\,\tilde\phi = - i\,m\,{B}, \nonumber\\
&& s_w\,\psi = e\,B\,\gamma_5\,\psi - e\,{\cal B}\,\psi, \qquad s_w\,\bar\psi 
= e\,B\,\bar\psi\,\gamma_5 + e\,{\cal B}\,\bar\psi, \nonumber\\
&& s_w\,C = s_w\,\bar C = s_w\,B = s_w\,{\cal B} = 0, \qquad s_w^2 \ne 0,
\end{eqnarray}
\begin{eqnarray}
&& s_{\bar w}\,A_\mu =  i\,\varepsilon_{\mu\nu}\,\partial^\nu\,B + i\,\partial_\mu\,{\cal B}, \qquad 
s_{\bar w}\,\phi =  i\,m\,{\cal B}, \qquad s_{\bar w}\,\tilde\phi =  i\,m\,{B}, \nonumber\\
&& s_{\bar w}\,\psi = - (e\,B\,\gamma_5\,\psi - e\,{\cal B}\,\psi), \qquad s_{\bar w}\,\bar\psi 
= - (e\,B\,\bar\psi\,\gamma_5 + e\,{\cal B}\,\bar\psi), \nonumber\\
&& s_{\bar w}\,C = s_{\bar w}\,\bar C = s_{\bar w}\,B = s_{\bar w}\,{\cal B} = 0, \qquad s_{\bar w}^2 \ne 0,
\end{eqnarray}
where we have considered {\it all} the basic and auxiliary fields of our chosen Lagrangian density
in Eq. (9). A close look at (20) and (21) demonstrates that the bosonic symmetry transformations
$s_w$ and $s_{\bar w}$ are {\it not} independent of each-other as they satisfy the following
\begin{eqnarray}
(s_w + s_{\bar w})\,\Sigma = 0, \qquad \Sigma = A_\mu, \phi, \tilde\phi, C, \bar C, B, {\cal B}, \psi, \bar\psi,
\end{eqnarray}
where $\Sigma$ is the generic field of the Lagrangian density (9). In other words, we have a {\it unique}
bosonic symmetry in the thoery. The sanctity of the above statement can be verified by the following 
transformation properties of the Lagrangian density (9):
\begin{eqnarray}
s_w\,{\cal L}_{\cal B} &=& \partial_\mu\,\Big[i\,m\,\varepsilon^{\mu\nu}\,\big\{\phi\,(\partial_\nu\,B) + m\,B\,A_\nu
\big\} + i\,m\,\tilde\phi\,\partial^\mu\,B + i\,{\cal B}\,\partial^\mu\,B - i\,B\,\partial^\mu\,{\cal B}\Big], \nonumber\\
s_{\bar w}\,{\cal L}_{\cal B} &=& - \partial_\mu\,\Big[i\,m\,\varepsilon^{\mu\nu}\,\big\{\phi\,(\partial_\nu\,B) + m\,B\,A_\nu
\big\} + i\,m\,\tilde\phi\,\partial^\mu\,B + i\,{\cal B}\,\partial^\mu\,B \nonumber\\ 
&-& i\,B\,\partial^\mu\,{\cal B}\Big]. 
\end{eqnarray}
This observation, once again, corroborates the fact that there is an existence of a {\it unique} [i.e. 
$(s_w + s_{\bar w})\,{\cal L}_{\cal B} = 0$] bosonic symmetry $(s_w)$ in our theory.

According to the Noether theorem, the above continuous bosonic symmetry leads to the derivation of 
the following conserved $(\partial_\mu\,J^\mu_{(w)} = 0)$ Noether current $(J^\mu_{(w)})$:
\begin{eqnarray}
J^\mu_{(w)} &=& i\,\varepsilon^{\mu\nu}\,\Big[{\cal B}\,\partial_\nu\,{\cal B} - B\,\partial_\nu\,B 
+ m\,\tilde\phi\,\partial_\nu\,{\cal B} - m\,\phi\,\partial_\nu\,B - m^2\,B\,A_\nu\Big] \nonumber\\ 
&+& i\,m\,B\,\partial^\mu\,\tilde\phi - i\,m\,{\cal B}\,\partial^\mu\,\phi + i\,m^2\,{\cal B}\,A^\mu
+ i\,e\,B\,\bar\psi\,\gamma^\mu\,\gamma_5\,\psi - i\,e\,{\cal B}\,\bar\psi\,\gamma^\mu\,\psi.
\end{eqnarray}
The conservation law $(\partial_\mu\,J^\mu_{(w)} = 0)$ can be proven by using the beauty and strength of the EL-EOMs 
that have been derived in Eq. (15) from the Lagrangian density (9). The bosonic
conserved charge $Q_w = \int d\,x\, J^0_{(w)}$ (with $\gamma_0\,\gamma_5 = \gamma_1$) is as follows:
\begin{eqnarray}
Q_w &=& \int d\,x\,\Big[i\,B\,\partial_1\,B - i\,{\cal B}\,\partial_1\,{\cal B} + i\,m\,B\,\dot{\tilde\phi}
- i\,m\,{\cal B}\,\dot\phi + i\,m\,\phi\,\partial_1\,B - i\,m\,\tilde\phi\,\partial_1\,{\cal B} \nonumber\\
&+& i\,m^2\,B\,A_1 + i\,m^2\,{\cal B}\,A_0 + i\,e\,B\,\bar\psi\,\gamma_1\,\psi 
- i\,e\,{\cal B}\,\bar\psi\,\gamma_0\,\psi\Big].
\end{eqnarray}
It is straightforward to check that the above conserved charge $(Q_w)$ is the generator of the continuous 
and infinitesimal {\it bosonic} symmetry transformations in Eq. (20).

We have a ghost-scale symmetry in our theory where {\it only} the fermionic $(C^2 = {\bar C}^2 = 0, \,
C\,\bar C + \bar C\,C = 0)$ (anti-)ghost fields $(\bar C)\,C$ transform continuously by a scale factor:
\begin{eqnarray}
C \rightarrow C = e^\Omega\,C, \qquad \bar C \rightarrow \bar C = e^{- \Omega}\,\bar C, \qquad 
\Sigma \rightarrow e^0\,\Sigma,
\end{eqnarray} 
In the above, the field $\Sigma = A_\mu, \phi, \tilde\phi, B, {\cal B}, \psi, \bar\psi$ stands for the generic field of
${\cal L}_{\cal B}$ [{\it without} the fermionic (anti-)ghost fields] and $\Omega$ in the exponentials of
Eq. (26) is the global (i.e. spacetime independent) scale factor. The signs in the exponents of $C$ and
$\bar C$ denote the ghost numbers $(\pm 1)$, respectively. The infinitesimal versions of 
(26) are  
\begin{eqnarray}
s_g\,C = + C, \qquad s_g\,\bar C = - \bar C, \qquad s_g\,\Sigma = 0,
\end{eqnarray}
where, for the sake of brevity, we have taken $\Omega = 1$. The invariance of ${\cal L}_{\cal B}$
under the continuous and infinitesimal {\it global} ghost-scale symmetry transformations (27) leads to the
existence of the following Noether's conserved current:
\begin{eqnarray}
J^\mu_{(g)} = i\,\Big[\bar C\,(\partial^\mu\,C) - (\partial^\mu\,\bar C)\,C\Big].
\end{eqnarray}
The conservation law $(\partial_\mu\,J^\mu_{(g)} = 0)$ can be proven by using the EL-EOMs that have 
been quoted in Eq. (15). The conserved charge $Q_g = \int d\,x\,J^0_{(g)}$ is as follows:
\begin{eqnarray}
Q_{g} = i\,\int d\,x\,\big(\bar C\,\dot C - \dot{\bar C}\,C\big).
\end{eqnarray}
The subscripts $g$ in Eqs. (27), (28) and (29) denote the infinitesimal transformations, conserved current
and conserved charge corresponding to the presence of an infinitesimal and continuous {\it global}
ghost-scale symmetry transformation in our theory.

We end this section with the following key comments. First of all, we note that the (anti-)ghost fields
$(\bar C)\,C$ do {\it not} transform under the bosonic symmetry transformations [cf. Eq. (20)]. Second, out of 
the {\it four} fermionic symmetries [i.e. $s_{(a)b}$ and $s_{(a)d}$] of our theory, it can be checked that 
the anticommutators: $\{s_b, s_{ab}\},\, \{s_d, s_{ad}\},\, \{s_b, s_{ad}\},\, \{s_d, s_{ab}\}$ turn out to be
{\it absolutely} zero. Third, only two {\it non-trivial} anticommutators (i.e. $s_w = \{s_b, s_d\}, \, s_{\bar w} = \{s_{ad}, 
s_{ab}\}$) exist. However, only {\it one} of them turns out to be {\it independent} which defines a {\it unique} bosonic
symmetry transformation $(s_w)$ for our theory [cf. Eq. (20)] because of our observation: $s_w + s_{\bar w} = 0$
[cf. Eqs. (20), (21), (23)]. Fourth, under the ghost-scale symmetry transformations, only the (anti-)ghost fields
transform by a {\it global} scale factor. However, all the rest of the fields do {\it not} transform at all. 
Finally, we observe the following interesting relationships, namely;
\begin{eqnarray}
s_w\,Q_r &=& - i\,\big[Q_r, Q_w\big] = 0, \qquad \qquad \quad r = b, ab, d, ad, g, w, \nonumber\\
s_g\,Q_s &=& - i\,\big[Q_s, Q_g\big] = + Q_s, \qquad \qquad s = b, ad, \nonumber\\
s_g\,Q_t &=& - i\,\big[Q_t, Q_g\big] = - Q_t, \qquad \qquad t = d, ab,
\end{eqnarray}   
which demonstrate that the conserved charge $Q_w$ is the Casimir operator for the whole algebra constructed 
with the conserved
charges $Q_r (r = b, ab, d, ad, g, w)$. Furthermore, we note that the ghost number for the conserved charges
$Q_b$ and $Q_{ad}$ is $+ 1$ and the {\it same} for the conserved charges $Q_d$ and $Q_{ab}$ is $(- 1)$. We shall
corroborate these statements very clearly in our forthcoming sections 5 and 6, respectively, where these
observations will be shown to be {\it important} in the proof of our present theory to be a model for Hodge theory.

\section{Algebraic Structures: Symmetries and Charges}

In this section, we assimilate {\it all} the continuous symmetry transformations for our 2D theory and 
point out their algebraic structure where they {\it act} like operators. In other words, when we say that 
$\{s_b, s_{ab}\} = 0$, it implies that we have: $\{s_b, s_{ab}\}\,\Sigma = 0$ where $\Sigma = A_\mu, C, \bar C, 
B, {\cal B}, \psi, \bar\psi, \phi, \tilde\phi$ is the generic field. It is straightforward to note that we 
obtain the following extended BRST-algebra that is obeyed by the infinitesimal and continuous symmetry 
transformation operators $s_r \,(r = b, ab, d, ad, g, w)$ of our theory, namely:   
\begin{eqnarray}
&& \{s_b, s_{ab}\} = 0, \qquad \{s_d, s_{ad}\} = 0, \qquad s_{(a)b}^2 = 0, \qquad s_{(a)d}^2 = 0, \nonumber\\
&& \{s_b, s_{ad}\} = 0, \qquad \{s_d, s_{ab}\} = 0, \qquad \{s_d, s_b\} = - \{s_{ab}, s_{ad}\} = s_w, \nonumber\\ 
&& [s_w, s_r] = 0, \qquad r = b, ab, d, ad, g, w, \nonumber\\
&& [s_g, s_b] = + s_b, \qquad [s_g, s_{ad}] = + s_{ad}, \qquad [s_g, s_{ab}] = - s_{ab}, \qquad [s_g, s_d] = - s_d, 
\end{eqnarray}
which shows that the bosonic symmetry transformation $(s_w)$ commute with {\it all} the {\it other} continuous
and infinitesimal symmetry transformations $s_r\,(r = b, ab, d, ad, g)$ of our 2D theory. On the other hand, the commutator
of $s_g$ with $s_r \,(r = b, ad)$ produces the transformations $s_r$ with {\it positive} sign (which is
just the {\it opposite} of the commutators of $s_g$ with $s_r\,(r = d, ab)$ where there is a {\it negative} sign
on the r.h.s.).

There is a very beautiful relationship between the fermionic symmetry transformations and the discrete
symmetry transformations (10) of our 2D theory. We very clearly observe the validity of the following
relationships:
\begin{eqnarray}   
s_d = \pm \, *\,s_b\,*, \qquad s_{ab} = \pm\,*\,s_{ad}\,*, 
\end{eqnarray}
where $*$, in the above, stands for the discrete symmetry transformations (10) of our theory. A close look
at (32) demonstrates that we have captured the mathematical relationship: $\delta = \pm\,*\,d\,*$ of differential
geometry in the terminology of the symmetry transformations where the {\it continuous} and {\it discrete} 
symmetry transformations are intertwined {\it together} and
play a very important and decisive role. Following the basic concepts behind the {\it duality} invariant theories
(see, e.g. [34] for details), we observe that the $(\pm)$ signs in the relationship: $s_d = \pm \, *\,s_b\,*$
are governed by the {\it double} operations of $*$ operator [i.e. the discrete symmetry transformations] on a
specific field. In other words, we note that for the generic field $\Psi$:
\begin{eqnarray}
&& *\,(*\,\Psi) \equiv - \Psi, \qquad \Psi = C, \bar C, A_\mu, B, {\cal B}, \phi, \tilde\phi, \nonumber\\
&& *\,(*\,\Psi) \equiv + \Psi, \qquad \Psi = \psi, \bar\psi.
\end{eqnarray}    
The above observations show that, {\it only} for the Dirac fields $(\psi, \bar\psi)$, we have the relationships:
$s_d\,\psi = +\,*\,s_b\,*\,\psi, \, s_d\,\bar\psi = +\,*\,s_b\,*\,\bar\psi$. On the contrary, for the {\it rest} of the
fields of our theory, we have: $s_d = - *\,s_b\,*$ where there is a {\it negative} sign on the r.h.s. We have provided more
discussions on it in our Appendix B.

According to the basic principle behind the Noether theorem, we know that the continuous symmetry transformations
and the Noether conserved charges (i.e. symmetry generators) are deeply connected with one-another.
Hence, it turns out that the algebra (31), obeyed by the symmetry transformation operators, is {\it also} respected and
replicated by the
conserved charges $Q_r\,(r = b, ab, d, ad, g, w)$. In other words, the algebra satisfied by the conserved charges
(which are responsible for the existence of the continuous and infinitesimal symmetry transformations in our theory) respect
the following:
\begin{eqnarray}
&& Q_{(a)b}^2 = 0, \qquad Q_{(a)d}^2 = 0, \qquad Q_w = \{Q_b, Q_d\} = - \,\{Q_{ad}, Q_{ab}\}, \nonumber\\
&& [Q_w, Q_r] = 0, \qquad r = b, ab, d, ad, g, w, \nonumber\\
&& \{Q_b, Q_{ab}\} = \{Q_d, Q_{ad}\} = \{Q_b, Q_{ad}\} = \{Q_{ab}, Q_d\} = 0, \nonumber\\
&& i\,[Q_g, Q_b] = + \,Q_b, \qquad i\,[Q_g, Q_{ad}] = + \,Q_{ad}, \nonumber\\
&& i\,[Q_g, Q_d] = -\, Q_d, \qquad i\,[Q_g, Q_{ab}] = -\, Q_{ab}.
\end{eqnarray}
It is crystal clear that the bosonic conserved charge $Q_w$, constructed from the off-shell nilpotent
(i.e. fermionic) conserved charges, is the Casimir operator for the whole algebra. A close and clear look
at (34) also shows, in a subtle manner, that the ghost numbers for the set of charges $(Q_b, Q_d, Q_w)$
are $(+ 1, - 1, 0)$, respectively. On the other hand, the set of charges $(Q_{ad}, Q_{ab}, Q_w)$ carries the
ghost numbers equal to $(+ 1, - 1, 0)$, respectively, too. Hence, we observe that there are two sets of conserved 
charges, in the algebra (34), which carry the ghost numbers $(+ 1, - 1, 0)$ in our theory. This observation would play 
a very important role in the {\it next} section where we shall establish the connection of the algebra (obeyed by the
{\it physical} conserved charges and continuous as well as discrete symmetries of our theory) 
{\it with} that of the algebra respected by the de Rham cohomological
operators of the differential geometry [8-12] which are purely {\it mathematical} in nature.

\section{Physical Symmetries, Conserved Charges and Cohomological Operators: Algebraic Connection}

In this section, we establish the precise connection between the discrete and continuous {\it physical}
symmetries (and conserved charges) of our field-theoretic model {\it and} the de Rham cohomological  
operators of differential geometry [8-12] at the level of algebra. In this context, first of all, 
we {\it recall} that the following explicit algebra [8-12] 
\begin{eqnarray}
&& d^2 = 0, \qquad \delta^2 = 0, \qquad \Delta = d\,\delta + \delta\,d = \{d, \delta\}, \nonumber\\
&& [\Delta, d] = 0, \qquad [\Delta, \delta] = 0, \qquad \Delta^2 \ne 0,
\end{eqnarray}
is satisfied by the de Rham cohomological operators $(d, \delta, \Delta)$ where the (co-)exterior derivatives
$(\delta)\,d$ obey: $\delta = \pm\,*\,d\,*$. Here the symbol $*$ stands for the Hodge duality operator on the
compact manifold {\it without} a boundary. Furthermore, we note that when $d$ operates on an arbitrary form $(f_n)$ of 
degree $n$, it raises the degree by {\it one} 
(i.e. $d\,f_n \sim f_{n + 1}$). On the other hand, we have:
$\delta\,f_n \sim f_{n - 1}$ which demonstrates that the operation of $\delta$ on a form decreases the degree of the
form by {\it one}. It is worthwhile to mention that here (i.e. $\delta\,f_n \sim f_{n - 1}$) the degree of the 
form is {\it non-zero} (i.e. $n = 1, 2, 3...$). The degree of a given form $(f_n)$ remains unchanged when 
it is operated upon by the Laplacian 
operator $\Delta$.

Against the backdrop of the above paragraph and equation (35), it is very important for us to provide the 
{\it physical} realizations of (35) and capture the key properties of the operations of the cohomological operators
on a given differential form $(f_n)$ of a non-zero degree $n$ in the language of the {\it physical} symmetries
and conserved charges (in the {\it quantum} Hilbert space). In this context, a clear and close look at the
algebra (34) demonstrates that there exists a deep connection between the cohomological operators and conserved
charges of our theory at the {\it algebraic} level. In fact, we find that the following two-to-one mapping
exists between the conserved charges and the cohomological operators:
\begin{eqnarray}  
&& (Q_b, Q_{ad}) \rightarrow d, \qquad (Q_d, Q_{ab}) \rightarrow \delta, \nonumber\\
&& Q_w = \{Q_b, Q_d\} \equiv - \{Q_{ad}, Q_{ab}\} \rightarrow \Delta.
\end{eqnarray}
As pointed out earlier, the absolute anticommutativity (i.e. $\{s_b, s_{ab}\} = 0,\, \{s_d, s_{ad}\} = 0$) properties
of the (anti-)BRST and (anti-)co-BRST symmetry transformations (and their corresponding conserved charges) imply
that {\it only} one of these symmetries (and corresponding conserved charge) can be identified with $d$ and $\delta$,
respectively. It turns out that, when $Q_b$ is identified with the exterior derivative $d$, the corresponding
dual-BRST charge $Q_d$ is identified with $\delta$. As a consequence, we find that the Laplacian operator
$\Delta$ is identified with $Q_w = \{Q_b, Q_d\}$ due to the fact that $\Delta = \{d, \delta\} = (d + \delta)^2$.
In exactly similar fashion, we find that when $Q_{ad}$ provides the physical realization of $d$, then, the co-exterior
derivative is identified with the anti-BRST charge $Q_{ab}$. Hence, the Laplacian operator turns out to be
$Q_w = - \{Q_{ad}, Q_{ab}\}$. The {\it negative} sign appears {\it here} due to the fact that $Q_w$ is {\it also} defined as:
$Q_w = \{Q_d, Q_b\}$ where there is a {\it plus} sign on the r.h.s. (in view of $s_w + s_{\bar w} = 0$).

As far as the change in the degree of the form, due to the operations of the cohomological operators is concerned,
we point out that {\it this} property can be realized in the quantum Hilbert space of states where we define (in terms
of the {\it conserved} ghost charge $Q_g$) any arbitrary
quantum state with a {\it non-zero} ghost number $n$ as:
\begin{eqnarray} 
i\,Q_g\,\mid \psi>_n = n\,\mid \psi>_n.
\end{eqnarray}
The algebra (34) now immediately leads to the following interesting and important observations as far as the  
ghost numbers of some {\it specific} states [constructed with $(Q_w,\, Q_b,\, Q_d)$ and $(Q_w,\, Q_{ad},\, Q_{ab})$] 
in the {\it quantum} Hilbert space are concerned, namely;
\begin{eqnarray}
i\,Q_g\,Q_b\,\mid \psi >_n &=& (n + 1)\,Q_b\,\mid \psi >_n, \nonumber\\
i\,Q_g\,Q_d\,\mid \psi >_n &=& (n - 1)\,Q_d\,\mid \psi >_n, \nonumber\\
i\,Q_g\,Q_w\,\mid \psi >_n &=& n\,Q_w\,\mid \psi >_n.
\end{eqnarray}
The above equations imply that the quantum states $Q_b\,\mid \psi >_n,\, Q_d\,\mid \psi >_n$ and
$Q_w\,\mid \psi >_n$ have the ghost numbers $(n + 1),\, (n - 1),\, n$, respectively. In other words, the ghost number
increases by {\it one} due to the operation of $Q_b$ {\it but} decreases by {\it one} due to the operation of $Q_d$.
On the contrary, the ghost number of a state remains intact due to the operation of $Q_w$ on it. In {\it exactly}
similar fashion, given equation (37) as an input, we have the following mathematical relationships, namely;
\begin{eqnarray}
i\,Q_g\,Q_{ad}\,\mid \psi >_n &=& (n + 1)\,Q_{ad}\,\mid \psi >_n, \nonumber\\
i\,Q_g\,Q_{ab}\,\mid \psi >_n &=& (n - 1)\,Q_{ab}\,\mid \psi >_n, \nonumber\\
i\,Q_g\,Q_w\,\mid \psi >_n &=& n\,Q_w\,\mid \psi >_n,
\end{eqnarray} 
which demonstrates that the ghost numbers are $(n +1),\, (n - 1)$ and $n$ for the states $Q_{ad}\,\mid \psi >_n,\,
Q_{ab}\,\mid \psi >_n$ and $Q_w\,\mid \psi >_n$, respectively. Hence, the operations of the set of charges 
$(Q_{ad},\, Q_{ab},\, Q_w)$, on a quantum state of the ghost number $n$, {\it are exactly} like the operations of the set of
de Rham cohomological operators $(d,\, \delta,\, \Delta)$ on a form of degree $n$. We can capture mathematically,
the observations made in Eqs. (38) and (39), as follows:
\begin{eqnarray} 
&& Q_b\,\mid \psi >_n \quad \sim \quad \mid \psi >_{n + 1}, \qquad Q_{ad}\,\mid \psi >_n \quad \sim 
\quad \mid \psi >_{n + 1}, \nonumber\\
&& Q_d\,\mid \psi >_n \quad \sim \quad \mid \psi >_{n - 1}, \qquad Q_{ab}\,\mid \psi >_n 
\quad \sim \quad \mid \psi >_{n - 1}, \nonumber\\
&& Q_w\,\mid \psi >_n \quad \sim \quad \mid \psi >_{n}, \qquad \quad Q_w\,\mid \psi >_n \quad \sim \quad \mid \psi >_{n}.
\end{eqnarray}
Thus, we observe that the changes in the degree of a given form, due to the operations of $(d,\, \delta,\, \Delta)$ are
{\it exactly} like the changes in the ghost numbers of a suitably chosen quantum state due to the operations 
of $(Q_b,\, Q_d,\, Q_w)$ 
and/or $(Q_{ad},\, Q_{ab},\, Q_w)$. In equations (37) to (40), we have denoted the ghost numbers of the states
by the {\it subscripts} on these states.

At this crucial juncture, we are in the position to capture the Hodge decomposition theorem of differential
geometry [8-12] in the language of the conserved charges of our 2D {\it interacting} theory in the quantum Hilbert 
space of states, namely;
\begin{eqnarray}
\mid \psi>_n &=& \mid \omega>_n \, + \, Q_b\,\mid \chi >_{n - 1} \, + \, Q_d\,\mid \lambda>_{n + 1} \nonumber\\
& \equiv & \mid \omega>_n \, + \, Q_{ad}\,\mid \sigma>_{n - 1} \, + \, Q_{ab}\,\mid \kappa>_{n + 1},  
\end{eqnarray}
where $\mid \psi>_n$ is any arbitrary quantum state in the Hilbert space with the ghost number $n$. For 
the sake of generality, we have chosen the non-null states $\mid \chi>_{n - 1},\, \mid \sigma>_{n - 1},\, \mid \lambda>_{n + 1}$
and $\mid \kappa>_{n + 1}$ in the Hodge decomposed state $\mid \psi>_n$ in terms of the {\it two} sets of charges 
$(Q_w,\, Q_b,\, Q_d)$ and $(Q_w,\, Q_{ad},\, Q_{ab})$, respectively. We note that the states $\mid \chi>_{n - 1}$ 
and $\mid \sigma>_{n - 1}$ are the BRST-exact and anti-co-BRST-exact, respectively. On the other hand, we have the 
states $\mid \lambda>_{n + 1}$ and $\mid \kappa>_{n + 1}$ which are co-BRST-exact and anti-BRST-exact, respectively. 
In the Hodge-decomposed state (41), we have the quantum state $\mid \omega>_n$ as the harmonic state which
is the {\it most} symmetric state in the whole theory because it is annihilated by all the {\it fermionic}
conserved charges. In other words, we have the following:
\begin{eqnarray}
Q_{(a)b}\,\mid \omega>_n = 0, \qquad Q_{(a)d}\,\mid \omega>_n = 0.
\end{eqnarray} 
We choose the {\it harmonic} state as the {\it physical} state of the theory which is annihilated 
by the conserved and nilpotent
BRST charge as well as the co-BRST charge. We have, purposely, {\it not} written the anti-BRST as well as the anti-co-BRST
charges because {\it these} do not lead to any {\it new} restrictions on the {\it physical} state of the theory. To be 
economical and precise, we demand that the physical state (i.e. $\mid phys>$) is the {\it one} which is annihilated by, at
least, the conserved BRST and co-BRST charges. Thus, the {\it physical} space of our 2D theory
is a {\it subspace} of the {\it total} quantum Hilbert
space which satisfies the {\it physicality} criteria:
\begin{eqnarray}
Q_b\, \mid phys> = 0, \qquad Q_d\, \mid phys> = 0.
\end{eqnarray}
In the next section, we shall discuss the {\it physical} consequences of (43), in detail, to demonstrate that 
our present study is {\it physically} interesting and useful.

\section{Harmonic State as Physical State: Consequences}

As pointed out in the previous section, it is the {\it harmonic} state in the Hodge decomposed state (of the {\it total} quantum 
Hilbert space) which is the most {\it symmetric} state of our {\it entire} theory. This state has to be annihilated
by the {\it conserved} and nilpotent BRST and co-BRST charges (which is the {\it most} economical and precise requirement).
In this context, we have the following (from the condition: $Q_b \mid phys> = 0$), namely;
\begin{eqnarray}
\Pi^0 \equiv B\mid phys> = 0 \quad &\Rightarrow& \quad - (\partial \cdot A +m\,\phi)\mid phys> = 0, \nonumber\\
\dot \Pi^0 \equiv \dot B \mid phys> = 0 \quad &\Rightarrow& \quad - \partial_0\,(\partial \cdot A +m\,\phi)\mid phys> 
= 0, \nonumber\\
&\equiv& (\vec\bigtriangledown  \cdot \vec E - m\,\Pi_\phi - e\,\psi^\dagger\,\psi)\mid phys> = 0, 
\end{eqnarray}
where we have used the concise expressions for charge from Eq. (6) and assumed that the (anti-)ghost fields are
{\it not} physical fields.  
The above restrictions on the {\it physical} state are consistent with Dirac's quantization condition
where the {\it physical} state of a {\it quantum} gauge theory {\it must} be annihilated by the {\it operator} form of the first-class
constraints (of the corresponding {\it classical} gauge theory). 
It is clear from the Lagrangian density (1) [and/or (9)] that we have:
$\Pi^0 = - (\partial \cdot A +m\,\phi) \equiv B$ and $\dot\Pi^0 = \dot B \equiv \vec\bigtriangledown  \cdot \vec E 
- m\,\Pi_\phi - e\,\psi^\dagger\,\psi$ where $\Pi_\phi = \dot\phi - m\,A_0$ is the canonical conjugate momentum
w.r.t. the {\it pure} scalar field $\phi$ (which is nothing but the St$\ddot u$ckelberg-field in the context of 
our 2D Proca theory).
We would like to point out that $\Pi^0$ is the conjugate momentum w.r.t. $A_0$ field and 
$\dot\Pi^0 = \dot B = \vec\bigtriangledown  \cdot \vec E - m\,\Pi_\phi - e\,\psi^\dagger\,\psi$ emerges out from the
EL-EOMs (5) and/or (15). In the 2D case, we have $\vec\bigtriangledown  \cdot \vec E = \partial_1\,E$ and, therefore,
the Gauss-law of divergence is: $\dot B = \partial_1\,E - m\,\Pi_\phi - e\,\psi^\dagger\,\psi$ which is derived from the
top entry of Eq. (15) with the inputs: $B = - (\partial \cdot A + m\,\phi)$ and ${\cal B} = E - m\,\tilde\phi$.
In more precise language, within the ambit of BRST formalism, we have the restrictions on the physical state [cf. Eq. (44)] 
as the {\it quantum} generalizations of the {\it first-class} constraints ($\Pi^0 \approx 0$ and
$\vec\bigtriangledown  \cdot \vec E - m\,\Pi_\phi - e\,\psi^\dagger\,\psi \approx 0$) on the {\it classical} gauge 
theory which become {\it operators} at the {\it quantum} level and they annihilate the {\it physical} state.

Let us focus on the proof of the conservation law $(\dot Q_b = 0)$ for the concise form of the nilpotent BRST charge
$(Q_b)$ that is quoted in Eq. (6). 
In this context, we note that the {\it conserved} current $J^{\mu}_{(b)}$ [cf. Eq. (4)] corresponding to the
infinitesimal and continuous BRST symmetry transformations (2) defines the BRST charge $(Q_b)$ in a straightforward
manner. A close look at the expression of $J^\mu_{(b)}$ in Eq. (4) shows that, for the proof of the conservation
law $\partial_\mu\,J^\mu_{(b)} = 0$, we invoke the validity of the conservation law $\partial_\mu\,J^\mu_{(m)} =
\partial_\mu\,[\bar\psi\,\gamma^\mu\,\psi] = 0$ where $J^\mu_{(m)}$ is the {\it polar-vector} current constructed with the
Dirac fields (which are the interacting {\it matter} fields in our theory). There is {\it no} problem for the proof 
of $\partial_\mu\,J^\mu_{(m)} = 0$ as the EL-EOM of Eq. (15)
lead to it.
It is straightforward to note that we have the following when we take a direct {\it time} derivative on (6), namely;
\begin{eqnarray}
\dot Q_b = \int d^{D -1}\,x\,\big[B\,\ddot C - \ddot B\,C\big].
\end{eqnarray}
Using the EL-EOMs from (5) and/or (15), we note that $(\Box + m^2)\,C = 0$ and $(\Box + m^2)\,B = 0$. In the 
derivation of the {\it latter}, we have used $(\Box + m^2)\,\phi = 0$ and the conservation of the {\it matter} 
current\footnote{The conservation of the matter current $(J^\mu_{(m)} = e\,\bar\psi\,\gamma^\mu\,\psi)$, constructed
with the Dirac fields $(\bar\psi,\, \psi)$, is sacrosanct because this leads to the 
conservation of the electric charge which is
{\it universal} and true at the {\it classical} as well as {\it quantum} level (in any {\it physically} allowed
process).}
[i.e. $\partial_\mu\,J^\mu_{(m)} \equiv e\,\partial_\mu\,(\bar\psi\,\gamma^\mu\,\psi) = 0$] which is valid due to the
local gauge and/or BRST symmetry invariance in our theory. As a consequence of the above EL-EOMs, we have: 
$\ddot C = \vec\bigtriangledown^2\,C - m^2\,C$
and $\ddot B = \vec\bigtriangledown^2\,B - m^2\,B$. Substitutions of these into (45) lead to the following
explicit form of the conservation law, namely;
\begin{eqnarray}
\dot Q_b = \int d^{D -1}\,x\,\big[\vec\bigtriangledown \cdot (B\,\vec\bigtriangledown\,C - C\,\vec\bigtriangledown\,B)\big] 
\quad \longrightarrow \quad 0.
\end{eqnarray}
Thus, we find that the conservation of BRST charge is {\it true} in any arbitrary dimension of spacetime
because we see that $\dot Q_b \rightarrow 0$ due to Gauss's divergence theorem. For 2D case, 
we have $\dot Q_b = \int d\,x\,\big[\partial_1\,(B\,\partial_1\,C - 
C\,\partial_1\,B)\big] \rightarrow 0$. We conclude that the {\it physicality} criterion with the nilpotent
and conserved BRST charge: $Q_b\,\mid phys> = 0$
leads to the annihilation
of the {\it physical} state by the operator forms of the {\it primary} constraint $(\Pi^0 \approx  0)$ and 
the secondary constraint: $(\vec\bigtriangledown  \cdot \vec E - m\,\Pi_\phi - e\,\psi^\dagger\,\psi \approx 0)$ 
on our 2D St$\ddot u$ckelberg's modified Proca theory.
It is clear that {\it both} these constraints are {\it first-class}.

Due to the presence of a set of discrete symmetry transformations (10) and our observations in Secs. 5 and 6, it is
obvious that our present 2D theory is (i) a {\it perfect} model of a {\it duality-invariant} theory [34], and (ii) an example
of Hodge theory. As far as the {\it duality} property is concerned, it will be noted that the discrete symmetry
transformation on the gauge field (i.e. $A_\mu \rightarrow \mp\,i\,\varepsilon_{\mu\nu}\,A^\nu$) owes its
origin to the {\it self-duality} (see, e.g. [5, 7] for details) condition [i.e. $*\,A^{(1)} = *\,(d\,x^\mu\,A_\mu)$]
on the 1-form $(A^{(1)} = d\,x^\mu\,A_\mu)$ gauge field $A_\mu$ where the $*$ is the Hodge duality operation.
Rest of the transformations in (10) are {\it consistent} with {\it this} self-duality requirement. 
For a perfect {\it duality} invariant {\it gauge} theory, it is very {\it sacrosanct}
requirement that the {\it dual} to the {\it first-class} constraints [cf. Eq. (44)] {\it must} also annihilate the physical state
of the theory. In this context, it is interesting 
to point out that $(E - m\,\tilde\phi)$ is the {\it dual} of the gauge-fixing term $(\partial \cdot A + m\,\phi)$ due to
the discrete symmetry transformations (10). Similarly, we note that $\partial_0\,(E - m\,\tilde\phi)$ is {\it also}
dual to $[\partial_0\,(\partial \cdot A + m\,\phi)]$. The {\it dual} restrictions emerge out from the co-BRST charge
$Q_d^{(1)}$ as discussed below. We lay emphasis on the fact that operator forms of the first-class constraints
and their {\it dual} must annihilate the {\it physical} state at the level of {\it tree} and loop diagrams.
In other words, these constraints/restrictions on the physical state are {\it universal} for a perfect {\it duality} invariant gauge theory
and they must be respected at the {\it classical} as well as {\it quantum} level.

Against the backdrop of the {\it above}, the {\it physicality} criterion: 
$Q_d^{(1)}\,\mid phys> = 0$ imposes {\it exactly} the {\it conditions} on the physical 
state due to the {\it duality} considerations, namely;
\begin{eqnarray}  
{\cal B}\,\mid phys> = 0 \quad &\Rightarrow& \quad (E - m\,\tilde\phi)\,\mid phys> = 0, \nonumber\\
\dot{\cal B}\,\mid phys> = 0 \quad &\Rightarrow& \quad \partial_0\,(E - m\,\tilde\phi)\,\mid phys> = 0 \nonumber\\
&\equiv & (\partial_1\,\Pi^0 + m\,\Pi_{\tilde\phi} + e\,\psi^{\dagger}\,\gamma_5\,\psi + m\,\partial_1\,\phi - m^2\,A_1)
\,\mid phys> = 0,
\end{eqnarray}
where we have used the {\it concise} expression for the conserved co-BRST charge $Q_d^{(1)}$ [cf. Eq. (17)]
and EL-EOMs from Eq. (15) in the derivation of $\partial_0\,(E - m\,\tilde\phi)$. We have {\it also} used the 
expressions: $\Pi^0 = B,\, \Pi_{\tilde\phi} = - \,\dot{\tilde\phi},\, {\cal B} = E - m\,\tilde\phi$.  
Let us now concentrate on the {\it physical} consequence that
emerges out from the {\it top} entry of (47). A close and careful look at the 2D (anti-)BRST and (anti-)co-BRST
invariant Lagrangian density (9) demonstrates that the pseudo-scalar field $\tilde\phi$ has appeared in the
theory with a {\it negative} kinetic term. Such fields have been christened as the ``phantom" fields in the realm of
cosmology and they have been found to be useful in the context of cyclic, bouncing and self-accelerated models of 
Universe (see e.g. [26-30] for details). In a nut-shell, such {\it exotic} fields (with negative kinetic terms)
are {\it unphysical} in the sense that they have {\it not} yet been {\it detected} by the experiments. Thus, just {\it like} the
(anti-)ghost fields, such kinds of fields (e.g. the pseudo-scalar field $\tilde\phi$) do not {\it impose} any 
restriction on the physical state of a quantum gauge theory. This leads us to draw the conclusion that {\it physically},
the restrictions in (47), imply the following (from the physicality requirement $Q_d^{(1)}\,\mid phys> = 0$), namely;
\begin{eqnarray}       
{\cal B}\,\mid phys> = 0 \quad &\Rightarrow& \quad (E - m\,\tilde\phi)\,\mid phys> = 0 \qquad  \Rightarrow \quad 
E\,\mid phys> = 0, \nonumber\\
\dot{\cal B}\,\mid phys> = 0 \quad &\Rightarrow& \quad \partial_0\,(E - m\,\tilde\phi)\,\mid phys> = 0 \quad 
\Rightarrow \quad \dot E\,\mid phys> = 0.
\end{eqnarray}
At this crucial juncture, we recall that the electric field $E = - \frac{1}{2}\,\varepsilon^{\mu\nu}\,F_{\mu\nu} \equiv 
- \varepsilon^{\mu\nu}\,\partial_\mu\,A_\nu$ is {\it also} the {\it anomaly} term in 2D because $\partial_\mu\,(\bar\psi\,
\gamma^\mu\,\gamma_5\,\psi) = - \frac{\alpha}{2}\,\varepsilon^{\mu\nu}\,F_{\mu\nu} = \alpha\,E$ where $\alpha$
is a constant factor (that is {\it not} very important for our {\it present} discussion). Thus, the anomaly term and the requirement of 
its time-evolution invariance {\it must} annihilate the physical state.

We have the conservation $({\dot Q}_b = 0)$ of the BRST charge due to the conservation $(\partial_\mu\,J^\mu_{(b)} = 0)$
of the BRST invariant $(s_b\,J^\mu_{(b)} = 0)$ Noether current $(J^\mu_{(b)})$ which, in a {\it precise} language, 
implies the following in terms of the {\it physical} state (see, e.g. [36] for details)
\begin{eqnarray} 
<phys\mid \partial_\mu\,J^\mu_{(b)}\mid phys> = 0 \quad \Rightarrow \quad <phys\mid {\dot Q}_{(b)}\mid phys> = 0.
\end{eqnarray}
Against this as a backdrop, we now concentrate on the proof of the conservation $({\dot Q}_d^{(1)} = 0)$ of the 
concise form of the co-BRST charge in Eq. (17). It is straightforward to note that we have the following explicit
expression for $({\dot Q}_d^{(1)} \equiv d\,Q_d^{(1)}/d\,t)$, namely;
\begin{eqnarray}  
{\dot Q}_d^{(1)} = \int d\,x\,\big[{\cal B}\,\ddot{\bar C} - \ddot{\cal B}\,\bar C\big],
\end{eqnarray}
where we have applied {\it directly} the time derivative on (17). We have the EL-EOMs: $(\Box + m^2)\,\bar C = 0$
and $(\Box + m^2)\,{\cal B} = 0$. The {\it latter} is {\it true} at the {\it classical} level (where there are no loop-diagrams
and the {\it massless} limit of the Dirac fermions is taken into account). To be precise, we lay emphasis on the
fact that we find that $(\Box + m^2)\,{\cal B} = e\,\partial_\mu\,\big[\bar\psi\,\gamma^\mu\,\gamma_5\,\psi\big]$
where the r.h.s. is {\it precisely} equal to zero in the {\it classical} limit as stated {\it above}. When we consider the
triangle Feynman diagram, we obtain the ABJ anomaly term on the r.h.s. where 
$\partial_\mu\,(\bar\psi\,\gamma^\mu\,\gamma_5\,\psi) = \alpha\,E$ for the 2D theory (with $\alpha$ as a 
constant factor). Thus, we find that Eq. (50) reduces to the following explicit form:
\begin{eqnarray}
{\dot Q}_d^{(1)} &=& \int d\,x\,\big[{\cal B}\,(\partial_1^2\,\bar C - m^2\,\bar C) - (\partial_1^2\,{\cal B} 
- m^2\,{\cal B} + \alpha\,E)\,\bar C\big] \nonumber\\
& \equiv & \int d\,x\,\Big[\partial_1\big\{{\cal B}\,\partial_1\,\bar C - (\partial_1\,{\cal B})\,\bar C\big\}\Big] 
- \alpha\,\int d\,x\,E\,\bar C,
\end{eqnarray}  
where we have used $\ddot{\bar C} = \partial_1\,\bar C - m^2\,\bar C,\, \ddot{\cal B} = \partial_1^2\,{\cal B} 
- m^2\,{\cal B} + \alpha\,E$. It is straightforward to note that the {\it first} term of (51) goes to zero as
$x \rightarrow \pm\,\infty$ due to Gauss's divergence theorem. However, the {\it second} term (due to 2D anomaly)
remains intact. In view of (49), we note that: 
$<phys\mid \,\partial_\mu\,J^\mu_{(d)}\,\mid phys> \, \Rightarrow 
\,<phys\mid \,{\dot Q}_d^{(1)}\,\mid phys>$. 
However, as pointed out earlier, our theory is a {\it perfect} model of {\it duality} invariant theory [34] because 
$A_\mu \rightarrow \mp\,i\,\varepsilon_{\mu\nu}\,A^\nu$ [cf. Eq. (10)] arises due to the {\it self-duality}
restriction (see, e.g. [5, 7] for details). Thus, the
restrictions (48) are valid at the tree-level as well as loop-level Feynman diagrams\footnote{It is the very elegant
interplay between the co-BRST symmetry transformations (11) and the discrete symmetry transformations (10) that ensures
the {\it conservation} of the co-BRST charge at the {\it tree} as well as the {\it loop-level} diagrams (when we choose the {\it physical}
state as the {\it harmonic} state).}. At this crucial stage,
we find the following explicit expressions (see, e.g. [36] for details) 
\begin{eqnarray}
<phys\mid \partial_\mu\,J^\mu_{(d)}\mid phys> &=& \alpha\,<phys\mid E \mid phys> = 0 \quad \Rightarrow \nonumber\\
<phys\mid {\dot Q}_{d}^{(1)}\mid phys> &=& \alpha\,\int <phys\mid E\,\bar C \mid phys>\,dx = 0.
\end{eqnarray}
The r.h.s. of the above expressions are zero due to the fact that, in view of Eq. (48), we have:  $E\,\mid phys> = 0$.
In the proof of the conservation of co-BRST charge, we note that the anti-ghost field $\bar C$ does {\it not}
impose any restriction on the {\it physical} state.
Hence, in 2D, the {\it anomaly} term is {\it trivial} in the sense that the {\it vector}
and {\it axial-vector} currents (and corresponding Noether charges) are conserved {\it together} as far as the 
{\it physical} state of our 2D {\it interacting} theory is concerned. This is the {\it basic} reason behind the consistency 
and unitarity of the 2D anomalous Abelian 1-form gauge theories (see, e.g. [37-39] for details).

\section{Summary and Outlook}

In our present investigation, we have demonstrated that an {\it interacting} 2D {\it modified} version of Proca theory
(in {\it interaction} with the Dirac fields) represents a tractable field-theoretic model 
for the Hodge theory where the discrete and continuous
symmetries of {\it this} theory provide the {\it physical} realizations of the de Rham cohomological operators of differential
geometry at the {\it algebraic} level. Whereas the infinitesimal and continuous symmetries (and corresponding conserved
charges) of our present 2D theory provide the physical realizations of the cohomological operators, the discrete symmetry
transformations correspond to the Hodge duality $*$ operation of differential geometry\footnote{Under the discrete
symmetry transformations (10), the kinetic term $(\bar\psi\,i\,\gamma^\mu\,\partial_\mu\,\psi)$ and the mass
term $(- m\,\bar\psi\,\psi)$ of the Dirac fields remain {\it trivially} invariant. The interaction term
$(-\, e\,\bar\psi\,\gamma^\mu\,A_\mu\,\psi)$
is {\it also} found to respect the discrete symmetry transformations (10) [cf. Appendix C for details].}. The ghost number 
considerations of the quantum states in the {\it total} Hilbert space, in terms of the conserved charges, have been
able to provide the {\it physical} analogue of the {\it operations} of cohomological operators on the differential form
of a given degree and the {\it changes} (in the degree) that ensue.

For our 2D {\it interacting} theory, we have been able to demonstrate that (i) the nilpotent (anti-)BRST symmetry transformations
(and conserved charges) exist [cf. Eqs. (2), (6)] corresponding to the {\it local} gauge symmetry in the theory
which is generated by the first-class constraints, and (ii) the nilpotent (anti-)co-BRST symmetries (and corresponding conserved
charges) are present [cf. Eqs. (11), (17)] in the theory due to the existence of the local {\it chiral} symmetry (in the
{\it massless} limit of the Dirac fields). The {\it latter} symmetry exists due to the presence of a {\it duality} 
symmetry in the theory [cf. Eq. (10)]. As is well-known, there is presence of the ABJ anomaly whenever the local 
{\it gauge} and {\it chiral} symmetries co-exist {\it together} in a theory with Dirac fields (in {\it interaction} with a
gauge field). However, at the {\it classical} level (in the {\it massless} limit of the Dirac fields), this {\it anomaly}
term is zero (at the {\it tree} level Feynman diagrams). For the QED in 2D, the anomaly term is nothing but the
electric field (i.e. $\frac{\alpha}{2}\,\varepsilon^{\mu\nu}\,F_{\mu\nu} = \alpha\,E$ with $\alpha$ as a constant factor)
which appears in the conservation law of the {\it chiral} current (i.e. $\partial_\mu\,J^\mu_{(5)} = \partial_\mu\,
(\bar\psi\,\gamma^\mu\,\gamma_5\,\psi) = \alpha\,E$).

As per Dirac's quantization condition, a {\it physical} state of the {\it total} quantum Hilbert space must be annihilated
by the {\it operator} form of the {\it constraints} of a given {\it classical} theory. In our present 2D theory, {\it this} 
key feature is incorporated in the requirement that the {\it physical} state must be annihilated by the {\it conserved}
and nilpotent BRST charge [cf. Eq. (44)]. However, if a theory is perfectly {\it duality} invariant, the {\it dual} version of
these constraints must {\it also} annihilate the {\it physical} state at the quantum level [cf. Eq. (47)]. Our present 2D theory
is {\it not} only a {\it perfect} field-theoretic example of Hodge theory, 
it is {\it also} a beautiful model of the {\it duality}
invariant theory [34] due to the presence of the {\it discrete} symmetry transformations (10). This is the
reason that we have the restrictions (44) and (47) on the {\it physical} state of our theory which are {\it very}
sacrosanct and they {\it must} be respected at the {\it classical} as well as the {\it quantum} level. In other words,
the restrictions (44) and (47) are {\it true} at the {\it tree} as well as {\it loop-level} Feynman diagrams
for our 2D {\it interacting} theory.  This is precisely the reason that the 2D anomaly term becomes {\it trivial} and, as
a consequence, the {\it vector} ($J^\mu_{(b)} $) and axial-vector ($J^\mu_{(d)} $) currents are conserved {\it together} leading
to the conservation of the BRST and co-BRST charges at the {\it tree} as well as at the loop-level diagrams.

We end this section with a few crucial remarks. First of all, the {\it simultaneous} conservation of the BRST and 
co-BRST charges (and corresponding Noether currents) are true {\it only} in the case of 2D theory of QED
with Dirac fields. Second, in our earlier works [32, 33] on the dual-BRST symmetries ({\it without} mass term for the photon),
we have shown that the {\it physical} state is annihilated by the electric field $(E)$ and {\it its} time-evolution
$(\dot E)$ invariance (i.e. $E\,\mid phys> = 0,\, \dot E\,\mid phys> = 0$). However, in such theories, the infrared
divergence problem exists. In contrast, in our present theory of {\it massive} photon, 
such divergences do {\it not} exist (see, e.g. [36]). Third, if
QED with {\it massive} photon (along with a pseudo-scalar field) is taken into account, our conjecture is the 
observation that the {\it anomaly} term will be: $(E - m\,\tilde\phi)$. Fourth, our present analysis provides the
{\it basic} reason behind the consistency and unitarity of the 2D {\it anomalous} Abelian gauge theory with Dirac fields 
(see, e.g. [37-39]). Finally, the existence of the {\it negative}
kinetic term for the fields is {\it not} a problem in the modern-day cosmology where the cyclic, bouncing and self-
accelerated models of Universe require the existence of such kind of ``phantom" fields and particles (see, e.g. [26-30]). 
This pseudo-scalar field and an axial-vector  field (with negative kinectic terms) have also been
shown to provide a possible set of candidates for dark matter and dark energy [18-20]. 
Within the framework of (SUSY) quantum field theories, the St${\ddot u}$ckelberg-boson has been able to shed light on the
infrared problem in QED and it has been shown to play an important role
in providing a possible candidate for the {\it ultralight} dark matter. These issues have been recently discussed in   
a set of very interesting papers [40-42]. It will be nice future endeavor to apply BRST approach to these
(SUSY) field theoretic models.

\vskip 1cm

\noindent
{\bf Acknowledgments}\\

\noindent
Two (AT, AKR) of us gratefully acknowledge the financial support from Banaras Hindu University (BHU) under its 
{\it BHU-Fellowship} program. One (RPM) of us has learnt the {\it basics} of gauge theory, constraints, anomalous gauge theory,
etc., from Prof. R. Rajaraman who happens to be {\it one} of the most influential {\it mentors} of the QFT group at BHU. All the 
authors, very humbly and respectfully, dedicate their present work to {\it him}.

\vskip 1cm
\begin{center}
{\bf Appendix A: On Different Forms of Lagrangian Density}\\
\end{center}

\vskip 1cm

\noindent
The central theme of our present Appendix is to highlight that we have discussed a few different forms of
the (anti-)BRST and (anti-)co-BRST invariant Lagrangian densities (without the Dirac fields) in our earlier work [19]
{\it besides} the Lagrangian density (9) that has been taken for discussion in our present endeavor. For instance, we have taken
into account the following Lagrangian densities ({\it without} Dirac fields), namely;
\[
{\cal L}_{(B_1)} = \frac{1}{2}\,(E - m\,\tilde\phi) + m\,E\,\tilde\phi - \frac{1}{2}
\,\partial_\mu\,\tilde\phi\,\partial^\mu\,\tilde\phi + \frac{m^2}{2}\,A_\mu\,A^\mu + \frac{1}{2}
\,\partial_\mu\,\phi\,\partial^\mu\,\phi 
\]
\[
- m\,A_\mu\,\partial^\mu\,\phi - \frac{1}{2}\,(\partial \cdot A + m\,\phi)^2
- i\,\partial_\mu\,\bar C\,\partial^\mu\,C + i\,m^2\,\bar C\,C,
\eqno(A.1)
\]    
\[
{\cal L}_{(B_2)} = \frac{1}{2}\,(E + m\,\tilde\phi) - m\,E\,\tilde\phi - \frac{1}{2}
\,\partial_\mu\,\tilde\phi\,\partial^\mu\,\tilde\phi + \frac{m^2}{2}\,A_\mu\,A^\mu + \frac{1}{2}
\,\partial_\mu\,\phi\,\partial^\mu\,\phi 
\]
\[
+ m\,A_\mu\,\partial^\mu\,\phi - \frac{1}{2}\,(\partial \cdot A - m\,\phi)^2
- i\,\partial_\mu\,\bar C\,\partial^\mu\,C + i\,m^2\,\bar C\,C,
\eqno(A.2)
\]  
which are intimately connected with each-other by the discrete symmetry transformations: 
$A_\mu \rightarrow A_\mu,\, C \rightarrow C,\,
\bar C \rightarrow \bar C,\, \phi \rightarrow - \phi,\, \tilde\phi \rightarrow - \tilde\phi$. Thus, they do {\it not}
describe {\it two} different physical systems. It is straightforward to check that the following {\it on-shell}
nilpotent and absolutely anticommuting (anti-)BRST symmetry transformations $[s_{(a)b}]$
\[
s_{ab}\,A_\mu = \partial_\mu\,\bar C, \qquad s_{ab}\,C = i\,(\partial \cdot A + m\,\phi), \qquad s_{ab}\,\phi = m\,\bar C,
\]    
\[
s_{ab}\,(\bar C, E, \tilde\phi) = 0, \qquad s_{ab}\,(\partial \cdot A + m\,\phi) = (\Box + m^2)\,\bar C,
\]
\[
s_{b}\,A_\mu = \partial_\mu\,C, \qquad s_{b}\,\bar C = - i\,(\partial \cdot A + m\,\phi), \qquad s_{b}\,\phi = m\,C,
\]    
\[
s_{b}\,(C, E, \tilde\phi) = 0, \qquad s_{b}\,(\partial \cdot A + m\,\phi) = (\Box + m^2)\,C,
\eqno(A.3)
\]
transform the Lagrangian density ${\cal L}_{B_1}$ to a total spacetime derivative thereby rendering the
action integral $S = \int d\,x\,{\cal L}_{B_1}$ invariant. In exactly similar fashion, we can write the {\it on-shell}
nilpotent and absolutely anticommuting (anti-)BRST transformations for ${\cal L}_{B_2}$ 
from (A.3) by the replacements: $A_\mu \rightarrow A_\mu,\,
C \rightarrow C,\, \bar C \rightarrow \bar C,\, \phi \rightarrow - \phi,\, \tilde\phi \rightarrow - \tilde\phi$.

The Lagrangian density (A.1) also respects an {\it on-shell} nilpotent and absolutely anticommuting 
set of (anti-)co-BRST symmetry transformations $s_{(a)d}$, namely;
\[
s_{ad}\,A_\mu = - \varepsilon_{\mu\nu}\,\partial^\nu\,C, \qquad s_{ad}\,\bar C = i\,(E - m\,\tilde\phi), 
\qquad s_{ad}\,\tilde\phi = - m\,C,
\]    
\[
s_{ad}\,(C, \partial \cdot A + m\,\phi, \phi) = 0, \qquad s_{ad}\,(E - m\,\tilde\phi) = (\Box + m^2)\,C,
\]
\[
s_{d}\,A_\mu = - \varepsilon_{\mu\nu}\,\partial^\nu\,\bar C, \qquad s_{d}\,C = - i\,(E - m\,\tilde\phi), 
\qquad s_{d}\,\tilde\phi = - m\,\bar C,
\]    
\[
s_{d}\,(\bar C, \partial \cdot A + m\,\phi, \phi) = 0, \qquad s_{d}\,(E - m\,\tilde\phi) = (\Box + m^2)\,\bar C.
\eqno(A.4)
\]
It is straightforward to note that the Lagrangian density (A.1) transforms to a {\it total} spacetime derivative  
under the {\it on-shell} nilpotent (anti-)dual BRST symmetry transformations $s_{(a)d}$. As a consequence, the action
integral $S = \int d\,x\,{\cal L}_{B_1}$ remains invariant under $s_{(a)d}$. It goes without saying that the (anti-)co-BRST
symmetries for the Lagrangian density ${\cal L}_{B_2}$ can be obtained from (A.4) by the replacements: 
$A_\mu \rightarrow A_\mu,\, C \rightarrow C,\, \bar C \rightarrow \bar C,\, \phi \rightarrow - \phi,\, 
\tilde\phi \rightarrow - \tilde\phi$. As far as the {\it off-shell} nilpotent (anti-)BRST and (anti-)co-BRST 
{\it invariant} Lagrangian densities ({\it without} the Dirac fields) are concerned, we have the following [19]:
\[
{\cal L}_{\cal B} = {\cal B}\,(E - m\,\tilde\phi) - \frac{1}{2}\,{\cal B}^2 + m\,E\,\tilde\phi 
- \frac{1}{2}\,\partial_\mu\,\tilde\phi\,\partial^\mu\,\tilde\phi + \frac{m^2}{2}\,{A_\mu}{A^\mu} + \frac{1}{2}\,
\partial_\mu\,\phi\,\partial^\mu\,\phi 
\]
\[
- m\,A_\mu\,\partial^\mu\,\phi+ B\,(\partial \cdot A + m\,\phi) + \frac{1}{2}\,B^2 
- i\,\partial_\mu\,\bar C\,
\partial^\mu\,C + i\,m^2\,\bar C\,C,
\eqno(A.5)
\] 
\[
{\cal L}_{\cal {\bar B}} = {\cal {\bar B}}\,(E + m\,\tilde\phi) - \frac{1}{2}\,{\cal {\bar B}}^2 - m\,E\,\tilde\phi 
- \frac{1}{2}\,\partial_\mu\,\tilde\phi\,\partial^\mu\,\tilde\phi + \frac{m^2}{2}\,{A_\mu}{A^\mu} + \frac{1}{2}\,
\partial_\mu\,\phi\,\partial^\mu\,\phi
\]
\[
+ m\,A_\mu\,\partial^\mu\,\phi + {\bar B}\,(\partial \cdot A - m\,\phi) + \frac{1}{2}\,\bar B^2 
- i\,\partial_\mu\,\bar C\,
\partial^\mu\,C + i\,m^2\,\bar C\,C.
\eqno(A.6)
\] 
We point out that we have invoked {\it new} type of Nakanishi-Lautrup type auxiliary fields $({\cal {\bar B}}, \bar B)$
in the Lagrangian density (A.6) for the linearization of the kinetic and gauge-fixing terms. In our present endeavor,
we have concentrated {\it only} on the Lagrangian density (A.5) with the Dirac fields $(\bar\psi, \psi)$ 
in interaction with the {\it massive} gauge field $(A_\mu)$ for the sake of brevity. However, one can {\it also} 
focus on the Lagrangian density (A.6) for which the (anti-)BRST as well as the (anti-)co-BRST symmetry transformations 
have been written in our earlier work [19] on the modified 2D Proca theory {\it without} Dirac fields.
It can be {\it trivially} checked that (A.6) can {\it also} be generalized 
with Dirac fields $(\bar\psi, \psi)$ in interaction with massive gauge field $(A_\mu)$ as we have done in Eq. (9).

\vskip 1.5cm
\begin{center}
{\bf Appendix B: On the Physical Analogue of $\delta = \pm *\,d\,*$}\\
\end{center}

\vskip 1cm

\noindent
As discussed in Sec. 5, it is a well-known fact that the (co-)exterior derivatives $(\delta)\,d$ of
the differential geometry are connected to
each-other by the precise relationship: $\delta = \pm\,*\,d\,*$ where ($\pm$) signs are dictated 
by the dimensionality of the compact 
manifold without a boundary and the degree of the form that is invoked in the inner product [8-12].
The central purpose of our present Appendix is 
to provide the physical realization of ${\it this}$ relationship in the terminology of the discrete and 
continuous symmetry transformations of our 
present 2D theory. In this context, we take a few explicit examples to demonstrate that 
{\it all} the key ingredients of the algebraic 
relationship $\delta = \pm\,*\,d\,*$ can be {\it physically} realized. We note that, for the 
Dirac fields $(\bar\psi, \psi)$ and the gauge field $A_\mu$, we have the
following [34]
\[
*\,(*\,\psi) = + \psi, \qquad  *\,(*\,\bar\psi) = + \bar \psi, \qquad *\,(*\,A_\mu) = -\, A_\mu, 
\eqno(B.1)
\]
where $ *$ stands for the discrete symmetry transformations (10). First of all, we can verify that
the following relationships are correct, namely;
\[
s_d\,\psi = + *\,s_b\,*\,\psi, \qquad s_d\,\bar\psi = + *\,s_b\,*\,\bar\psi, \qquad s_d\,A_\mu = -\,* s_b\,* \,A_\mu,
\eqno(B.2)
\]
where the signs on the r.h.s. of (B.2) have been dictated by the signs in (B.1) as per the principles behind a 
duality invariant theory 
(see, e.g. [34] for details). In Eq. (B.2), the symbol $*$ stands, once again, for the 
discrete symmetry transformations (10). The 
nilpotent BRST symmetry transformations $s_b$ are given in Eq. (2) and the co-BRST symmetry 
transformations $s_d$ are quoted in Eq. (11).
It is crystal clear that (B.2) provides the physical realization of $\delta = \pm *\,d\,*$ where 
there is a mapping: $\delta \leftrightarrow s_d $
and $d \leftrightarrow s_b $. Furthermore, the Hodge duality $*$ operation of differential geometry 
is realized in terms of the discrete symmetry
transformations (10). We note that, exactly like Eq. (B.2), we have the validity of the following relationships
between $s_{ab}$ and $s_{ad}$, namely;
\[
s_{ab}\,\psi = + *\,s_{ad}\,*\,\psi, \qquad s_{ab}\,\bar\psi = + *\,s_{ad}\,*\,\bar\psi, \qquad 
s_{ab}\,A_\mu = -\,* s_{ad}\,* \,A_\mu,
\eqno(B.3)
\]
which demonstrate that we have the mappings between the {\it mathematical} quantities
$(\delta)\,d$ and the {\it physical} symmetry transformations
$(s_{ab})\,s_{ad}$ as: $\delta \leftrightarrow s_{ab},\, d \leftrightarrow s_{ad}$. In addition, 
the notation $*$ in (B.3) denotes, once again,
the discrete symmetry transformations (10).

We end this Appendix with the remarks that, it is the peculiarity of the specific two $(1+1)$-dimensional (2D) 
theory that we have {\it also} the validity of the {\it inverse} relationships corresponding to (B.2) and (B.3) as:
\[
s_{b}\,\psi = + *\,s_{d}\,*\,\psi, \qquad s_{b}\,\bar\psi = + *\,s_{d}\,*\,\bar\psi, \qquad 
s_{b}\,A_\mu = -\,* s_{d}\,* \,A_\mu, 
\]
\[
s_{ad}\,\psi = + *\,s_{ab}\,*\,\psi, \qquad s_{ad}\,\bar\psi = + *\,s_{ab}\,*\,\bar\psi, \qquad 
s_{ad}\,A_\mu = -\,* s_{ab}\,* \,A_\mu.
\eqno(B.4)
\]
Thus, it appears that we have {\it also} the sanctity of the mappings: $d \leftrightarrow (s_d,\, s_{ab})$ and
$\delta \leftrightarrow (s_b,\, s_{ad})$. However, this is {\it not} true as is evident from our discussions
on the ghost number considerations (cf. Sec. 6) in the {\it total} quantum Hilbert space in terms of conserved charges
corresponding to the continuous symmetry transformations generated by $s_b,\, s_{ab},\, s_d,\, s_{ad},\, s_w$ 
and $s_g$. It turns out that the ghost number increases by {\it one} when we apply the charges $(Q_b,\, Q_{ad})$ on 
a state with the ghost number $n$. On the contrary, the ghost number decreases by {\it one} due to the applications
of $(Q_d,\, Q_{ab})$ on a state with the ghost number $n$ in the {\it total} quantum Hilbert space of states.
Hence, the following identifications, namely;
\[
d \leftrightarrow (s_b, s_{ad}), \qquad \qquad \delta \leftrightarrow (s_d, s_{ab}),
\eqno(B.5)
\]  
are {\it true} as the conserved charges corresponding to the continuous symmetry transformations $(s_b, s_{ad})$
and $(s_d, s_{ab})$ capture the {\it real} and {\it sacrosanct} properties of $d$ and $\delta$. Furthermore, we note that
we have not taken other fields: $\phi,\, \tilde\phi,\, C,\, \bar C,\, B,\, {\cal B}$ of our theory in our present
discussion. However, as pointed out earlier, these fields will obey {\it exactly} the same kind of relationships as
has been respected by the $A_\mu$ field [cf. Eq. (33)].

\vskip 1cm
\begin{center}
{\bf Appendix C: On the Discrete Symmetries in Component Form}\\
\end{center}

\vskip 1cm

\noindent
We demonstrate here the {\it invariance} of the Dirac fields in {\it interaction} 
with a massive gauge field (i.e. $- e\,\bar\psi\,\gamma^\mu\,A_\mu\,\psi$) under the discrete symmetry 
transformations (10) where we take into account the component forms of the Dirac fields as well as the gauge field ($A_\mu$).
With the choices: $\gamma_0 = \sigma_1,\, \gamma^0 = \gamma_0$, we note that we have the following
for our 2D theory:
\[
\bar\psi = \psi^{\dagger}\,\gamma^0 = (\psi_{1}^{*},\,\psi_{2}^{*})\,\sigma_{1}  \equiv (\psi_{2}^{*},\,\psi_{1}^{*}).
\eqno (C.1)
\]
It is elementary to verify that the kinetic term for the fermionic Dirac Lagrangian density (i.e. 
$\bar\psi\,i\,\gamma^\mu\,\partial_\mu\,\psi$)
remains invariant under (10) because: $\psi \rightarrow \psi,\, \bar\psi \rightarrow \bar\psi$. 
Written in the component form, the interaction term
($-\,e\,\bar\psi\,\gamma^\mu\,A_\mu\,\psi$) in 2D is as follows:
\[
-\,e\,\bar\psi\,\gamma^\mu\,A_\mu\,\psi = -\,e\,\bar\psi\,\gamma^0\,A_0\,\psi  - \,e\,\bar\psi\,\gamma^1\,A_1\,\psi
\]
\[
\equiv -\,e\,\psi_{2}^{*}\,A_0\,\psi_2 -\,e\,\psi_{1}^{*}\,A_0\,\psi_1 +\,e\,
\psi_{2}^{*}\,A_1\,\psi_2 -\,e\,\psi_{1}^{*}\,A_1\,\psi_1.
\eqno (C.2)
\]
In the above, we have used (C.1) and taken into account $\gamma^0 = \gamma_0 = \sigma_1,\, \gamma^1 
= -\,\gamma_1 = -\,i\,\sigma_2 $.
We have also considered the column vectors: $\psi = (\psi_1, \psi_2)^{T}$ and $A_\mu = (A_1, A_2)^T$. It is
straightforward to check that the {\it latter} satisfies: $A_\mu \rightarrow \mp\,i\,\varepsilon_{\mu\nu}\,A^\nu$
{\it correctly} in the matrix form, too. In other words, we obtain: $A_0 \rightarrow \pm\,i\,A_1,\, 
A_1 \rightarrow \pm\,i\,A_0$
from the matrix forms of $\varepsilon_{\mu\nu}$ and $A_\mu$ where $\varepsilon_{\mu\nu}$ is a $2 \times 2$
matrix and $A_\mu$ is a $2 \times 1$ matrix.

To check the invariance of the interaction term ($-\,e\,\bar\psi\,\gamma^\mu\,A_\mu\,\psi$) under (10), first of all, we note 
that we have the following transformations under (10), namely;
\[
-\,e\,\bar\psi\,\gamma^\mu\,A_\mu\,\psi \quad \rightarrow \quad - (\pm\,i\, e \,\gamma_5)\,\bar\psi\,\gamma^\mu\,(\mp\,i\,
\varepsilon_{\mu\nu}\,A^{\nu})\,\psi 
\]
\[
= e\,\bar\psi\,\gamma_5\gamma^\mu\,\varepsilon_{\mu\nu}\,A^\nu\,\psi.
\eqno (C.3)
\]
In the above, we have used: $\gamma_5\,\bar\psi = -\, \bar\psi\,\gamma_5$ (as $\gamma_5\,\gamma_0 = - \gamma_0\,\gamma_5$).
Using the straightforward relationships: $\gamma^\mu\,\varepsilon_{\mu\nu} = - \gamma_5\,\gamma_\nu,\,
\gamma_5^2 = I$, we observe that the explicit form of (C.3) is nothing but the l.h.s. of (C.2) which proves
the point. However, we take into account the explicit expression (C.3) in the component form (with $\gamma^0 = \gamma_0,\,
\gamma^1 = - \gamma_1,\, A^0 = A_0,\, A^1 = - A_1$) as
\[
\gamma^\mu\,\varepsilon_{\mu\nu}\,A^\nu = \gamma^0\,\varepsilon_{01}\,A^1 + \gamma^1\,\varepsilon_{10}\,A^0 
= - \gamma_0\,A_1 + \gamma_1\,A_0,
\eqno(C.4)
\]  
to obtain the following form of the {\it interaction} term [cf. Eq. (C.3)]
\[
- e\,\bar\psi\,\gamma_5\,\gamma_0\,A_1\,\psi + e\,\bar\psi\,\gamma_5\,\gamma_1\,A_0\,\psi.
\eqno(C.5)
\] 
Substitutions of the component forms: $\bar\psi = (\psi_2^*,\, \psi_1^*),\, \psi = (\psi_1,\, \psi_2)^T,\, \gamma_5 = - 
\sigma_3,\, \gamma_0 = \sigma_1,\, \gamma_1 = i\,\sigma_2$ leads to the derivation of the following from (C.5):
\[
- e\,\bar\psi\,\gamma_5\,\gamma_0\,A_1\,\psi = e\,\psi_2^*\,A_1\,\psi_2 - e\,\psi_1^*\,A_1\,\psi_1,
\]
\[
e\,\bar\psi\,\gamma_5\,\gamma_1\,A_0\,\psi = - e\,\psi_2^*\,A_0\,\psi_2 - e\,\psi_1^*\,A_0\,\psi_1.\
\eqno(C.6)
\]
The sum of the above {\it two} terms [cf. Eqs. (C.5), (C.6)] is nothing but the Eq. (C.2). 
Thus, even in the component form, the {\it interaction} term remains {\it invariant} under (10). There is a {\it simpler}
way to verify the {\it same} thing. If we substitute $\gamma_5 = \gamma_0\,\gamma_1$ into (C.5) and
use $\gamma^0 = \gamma_0,\, \gamma^1 = - \gamma_1,\, \gamma_0^2 = I,\, \gamma_1^2 = - I$ and
$\gamma_0\,\gamma_1 = - \gamma_1\,\gamma_0$, we observe that Eq. (C.5) reduces to: $(- e\,\bar\psi\,\gamma^0\,A_0\,\psi 
- e\,\bar\psi\,\gamma^1\,A_1\,\psi)$ which was the starting point in Eq. (C.2).

\end{document}